\documentclass{aa} 
\usepackage{graphicx}
\usepackage{txfonts}
\usepackage{multicol}
\usepackage{multirow}
\usepackage{lscape}
\usepackage{adjustbox}
\usepackage{graphicx}
\usepackage{hyperref}   %call hyperlink pakage
\hypersetup{
    colorlinks=true,
   linkcolor=blue,
    filecolor=magenta,      
    urlcolor=cyan,
    }

\usepackage{amsmath,amssymb}
\usepackage{changepage}

\begin{document} 

   \title{Density calculations of NGC 3783 warm absorbers using a time-dependent photoionization model}

   %\subtitle{}

   \author{Chen Li
          \inst{1, 2}
          %\fnmsep\thanks
          \and
          Jelle S. Kaastra\inst{1, 2}
          \and
          Liyi Gu\inst{2, 1}
          \and
          Missagh Mehdipour\inst{3}
          }

   \institute{Leiden Observatory, Leiden University,
              P.O.Box 9513, 2300 RA Leiden, The Netherlands\\
              \email{cli@strw.leidenuniv.nl}
         \and
             SRON Netherlands Institute for Space Research,  Niels Bohrweg 4, 2333 CA Leiden, The Netherlands 
             %\email{c.ptolemy@hipparch.uheaven.space}
        \and
            Space Telescope Science Institute, 3700 San Martin Drive, Baltimore, MD 21218, USA
             }

   \date{}

% \abstract{}{}{}{}{} 
% 5 {} token are mandatory
 
  \abstract
  {
Outflowing wind as one type of AGN feedback, which involves noncollimated ionized winds prevalent in Seyfert-$1$ AGNs, impacts their host galaxy by carrying kinetic energy outwards.
However, the distance of the outflowing wind is poorly constrained due to a lack of direct imaging observations, which limits our understanding of their kinetic power and therefore makes the impact on the local galactic environment unclear.  
One potential approach involves a determination of the density of the ionized plasma, making it possible to derive the distance using the ionization parameter $\xi$, which can be measured based on the ionization state.
Here, by applying a new time-dependent photoionization model, \texttt{tpho}, in SPEX, we define a new approach, the \texttt{tpho}-delay method, to calculate/predict a detectable density range for warm absorbers of NGC $3783$.
The \texttt{tpho} model solves self-consistently the time-dependent ionic concentrations, which enables us to study delayed states of the plasma in detail.
%The source ionizing luminosity and the warm absorbers parameters obtained from previous data, together with the light curve derived from the power spectrum, were fed into the TPHO simulation. 
We show that it is crucial to model the non-equilibrium effects accurately for the delayed phase, where the non-equilibrium and equilibrium models diverge significantly.
Finally, we calculate the crossing time to consider the effect of the transverse motion of the outflow on the intrinsic luminosity variation.
It is expected that future spectroscopic observations with more sensitive instruments will provide more accurate constraints on the outflow density, and thereby on the feedback energetics.
}

  \titlerunning{TPHO simulation of NGC~3783 X-ray warm absorbers}
  \authorrunning{C. Li et al.}
  
  \maketitle  
%
%-------------------------------------------------------------------

%\tableofcontents      %the table of contents

\section{Introduction}\label{introduction}

Outflows from Active galactic nuclei (AGNs) have been found to carry significant kinetic energy, and thereby impact their host galaxy in a cosmic feedback mechanism, which usually shows imprint on the absorption spectrum in the UV and/or X-ray wavelength band \citep{Laha2021NatAs}.

With the assumption of spherical outflow \citep{Blustin2005A&A}, the mass outflow rate can be estimated via
\begin{equation}\label{equ:outflow mass rate}
\dot{M}_{\rm out} = 1.23 \, r^2 \, m_{p} \, n(r) \, v_{\rm out} \, C_{v}(r) \,  \Omega \, ,
\end{equation}
where the factor 1.23 takes into account the cosmic elemental abundances, $m_{p}$ is the mass of a proton, $n(r)$ is the density of the outflow at radius $r$, $C_{v}(r)$ is the volume filling factor as a function of distance and $\Omega$ is the solid angle subtended by the outflow.
Subsequently, the kinetic luminosity \( L_{\rm KE} = \frac{1}{2} \dot{M}_{\rm out} v_{\rm out}^{2} \) can be derived.
From the spectral fitting of absorption lines/edges in UV/optical and X-rays, one obtains the ionization parameter ($\xi$), the column density ($N_{\rm H}$) which rely on $C_{v}(r)$ together with $\Omega$, and the outflow velocity ($v_{\rm out}$).
However, the lack of spatial resolution in the inner parsec of the central engines, makes it difficult to view these flows directly, further impeding us to measure the impact of AGN feedback.

The ionization parameter (\citealp{Tarter1969ApJ}; \citealp{Krolik1981ApJ}) conveniently quantifies the ionization status of the outflows photoionized by the intense radiation from the accretion-driven central source, which is defined as:
\begin{equation}\label{equ:xi defination}
      \xi = \frac{L_{\rm ion}}{n_{\rm H} \times r^{2}} \,  ,
\end{equation}
where $L_{\rm ion}$ is the $1$-$1000 \rm \, Ryd$ (or $\rm 13.6 \, eV$–$\rm 13.6 \, keV$) band luminosity of the ionizing source, $n_{\rm H}$ the hydrogen number density of the ionized plasma, and $r$ the distance of the plasma to the ionizing source.
Therefore, the distances can be constrained indirectly through the measurements of ionization parameter, ionizing luminosity, and density.

Up to now, assessing the density of the outflow accurately remains challenging.
One approach is to use density sensitive lines from a spectral analysis.
The He-like triplet emission lines \citep{Porquet2010SSR} and the absorption lines from the metastable levels \citep{Arav2015AA}, among many other transitions, are considered to have a sensitivity to density.
\cite{Mao2017A&A} explored the density diagnostics through the use of the metastable absorption lines of Be-, B-, and C- like ions for the AGN outflows, and found that in the same isoelectronic sequence, different ions cover not only a wide range of ionization parameters but also an extensive density range.
Less ionized ions probe lower density and smaller ionization parameters within the same isonuclear sequence \citep{Mao2017A&A}.
The implementation of this method requires high-quality spectroscopic data of the outflows.

An alternative approach, spectral-timing analysis, mainly focuses on the variability of ionizing luminosity, involving investigating the changes in the outflows over time which can be a density-dependent response to the variation in the ionizing luminosity (\citealp{Kaastra2012A&A}; \citealp{Rogantini2022ApJ}).
How fast the ionized plasma responds is dependent on the comparison of recombination timescale and variability of the ionizing source.
Warm absorber (hereafter, WA) as one kind of outflowing wind, can be studied using this technique.
Through observations, it has been established that the response of WA to the continuum change provides valuable insight into the origin and acceleration mechanisms of WA \citep{Laha2021NatAs}. 
%Solely following the anti-correlated between recombination timescale and electron density, 
\cite{Ebrero2016A&A} estimated the lower limits on the density of WA of NGC $5548$ with long-term variability using the photoionization code Cloudy \citep{Ferland1998PASP}. 
Time-dependent photoionization modelling has been applied to Mrk $509$ \citep{Kaastra2012A&A} and NGC $4051$ \citep{Silva2016A&A}, respectively, to constrain the density of WAs.
It has been shown that the equilibrium model might become insufficient when it comes to detailed diagnostics of the outflows.
The non-equilibrium model is needed to interpret high-quality data including those obtained from future missions such as Athena (\citealp{Sadaula2022arXiv}; \citealp{Juranova2022MNRAS}).

Recently, \cite{Rogantini2022ApJ} build up a new time-dependent photoionization model, \texttt{tpho}, which performs a self-consistent calculation solving the full time-dependent ionization state for all ionic species and therefore allows investigating time-dependent ionic concentrations. 
The \texttt{tpho} model keeps the SED shape constant, while the luminosity is able to be changed by following the input light curve variation.
In this paper, we apply this method with a realistic example of an AGN.

NGC 3783, with redshift $z=0.009730$ \citep{Theureau1998A&AS}, hosts one of the most luminous local AGNs, with a bolometric AGN luminosity off log $L_{\rm AGN} \sim  44.5 \, \rm erg \, \rm s^{-1}$ at a distance of $38.5$ Mpc \citep{Davies2015ApJ} as well as hosting a supermassive black hole of $M_{\rm BH} = 3 \times 10^7 M_{\sun}$ (\citealp{Vestergaard2006ApJ}), and has been studied extensively, most notably for its ionised outflows (especially X-ray warm absorbers) and variability. 
%The X-ray-warm absorber was initially modelled by \cite{Netzer2003ApJ}, and now 
In the X-ray band, $10$ photoionized components have been found (\citealp{Mehdipour2017A&A}; \citealp{Mao2019A&A}).
%which can provide good WAs parameters to study time-dependent photoionization process.
The spectral energy distribution (hereafter, SED) measured by \cite{Mehdipour2017A&A} and power spectral density (hereafter, PSD) shape derived by \cite{Markowitz2005ApJ}, both in the unobscured state of NGC $3783$, give us realistic parameters to feed the \texttt{tpho} model. 
%In the obscured state, the X-ray obscurer in NGC 3783 varies on short timescales ranging between about one hour and ten hours \citep{DeMarco2020A&A}.
%Q1: ???Here we need note above sentence, it will be affect our calculation or not?

In this study, we execute time-dependent photoionization calculations using the \texttt{tpho} model \citep{Rogantini2022ApJ} in SPEX version $3.07.01$ \citep{kaastra2022} and adopt a realistic SED with variability as well as ionization parameters of $10$ WA components of NGC $3783$ based on previous observations (\citealp{Markowitz2005ApJ};  \citealp{Mehdipour2017A&A}; \citealp{Mao2019A&A}).
In Section \ref{sect:2}, we reproduce the measured SED and present the simulated light curve.
In Section \ref{sect:3}, we show \texttt{tpho} calculation results and execute a cross-correlation function to quantify the density-dependent lag, as well as compare \texttt{tpho} with \texttt{pion} (photoionization equilibrium model in SPEX).
In Section \ref{sect:4} and Section \ref{sect:5}, we discuss and conclude our current work.

%--------------------------------------------------------------------
\section{Method}\label{sect:2}

We first produce a representative SED based on previous measurements, and simulate a realistic light curve corresponding to the PSD shape for NGC $3783$.
%After 10 WAs was choose as target, we then execute TPHO calculation and therefore get the interpolation $\xi$ which varies as the luminosity of the input light curve going up/down (i.e. following the Eq. \ref{equ:xi defination}) but
We assume that the shape of the SED stays the same during the variations.
The \texttt{tpho} model calculates the evolution of ionic concentration for all available elements and all components of the WA.
Finally, we convert the ion concentration variation to the change of average charge states.

\subsection{Modelling the NGC 3783 SED}\label{subsect:sed}

We adopt the parameterized unobscured SED model of NGC $3783$ determined by \cite{Mehdipour2017A&A} for the unobscured state from $2000-2001$ mainly based on the archival data taken by XMM-Newton ($2000$ and $2001$) and Chandra HETGS ($2000$, $2001$).

%They used all archival XMM-Newton data (2000 and 2001) and Chandra HETGS data (2000, 2001, 2013) to produce a set of time-averaged spectra, and also provided better constrain of the 9 WAs photonized feature from these data (\citealp{Mao2019A&A}).
%All the archived unobscured data they used include X-ray and UV band and others (the part of references see \citealp{Kaspi2002ApJ}; \citealp{Blustin2002A&A}; \citealp{Behar2003ApJ}; \citealp{Scott2014ApJ}) to produce a set of time-averaged spectra. 

\begin{figure}[!tbp]
\centering
\includegraphics[width=\linewidth]{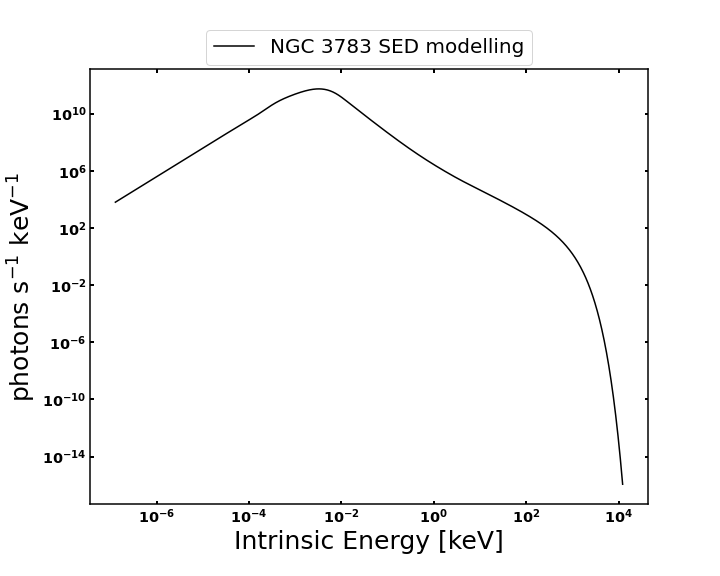}
\caption{SED of NGC 3783 used in this work, based on \cite{Mehdipour2017A&A} for the $2000-2001$ observation in an unobscured state.}
\label{fig:sed}
\end{figure}

This model 
%used by \cite{Mehdipour2017A&A} (see their paper's section 3 for the detail) 
has a SED consisting of an optical/UV thin disk component, an X-ray power-law continuum, a neutral X-ray reflection component, and a warm Comptonization component for the soft X-ray excess (Figure \ref{fig:sed}). 
%===========================

%---------------------------------------------------------------
\subsection{Source simulated light curve}\label{subsect:lc}
We use a doubly broken power spectral density (PSD, see Eq. \ref{equ:psd}) model to simulate the NGC $3783$ light curve \citep{Markowitz2005ApJ},

\begin{equation}\label{equ:psd}
P(f)=
    \begin{cases}
        A_l, & f \leq f_l, \\  
        A(f/f_h)^{-1},  & f_l < f \leq f_h, \\
        A(f/f_h)^{-\beta}, & f > f_h, 
    \end{cases}
\end{equation}
where $A_l$ is the PSD amplitude below the low-frequency break $f_l = 2 \times 10^{-7}$ Hz, here the PSD has zero slopes. 
$A = A_l(f_h/f_l)^{-1}$ is the PSD amplitude at the high-frequency break $f_h = 4 \times 10^{-6}$ Hz.
We adopt the best-fit Monte Carlo results from \cite{Markowitz2005ApJ} Table $2$ and set the slope value $\beta = 2.6$ as well as the power amplitude $A = 7200 \, \rm Hz^{-1}$ to simulate the source light curve with variability in the energy band of $0.2-12$ keV.
%The simulated light curve shows clear variability on $\rm 2.5 \times 10^{5} s - 5 \times 10^{6}s $.
The range between $\rm 2.5 \times 10^5 \, s$ and $\rm 5 \times 10^6 \, s$ becomes the typical timescale where high fluctuation occurs.

\begin{figure}[!tbp]
\centering
\includegraphics[width=\linewidth]{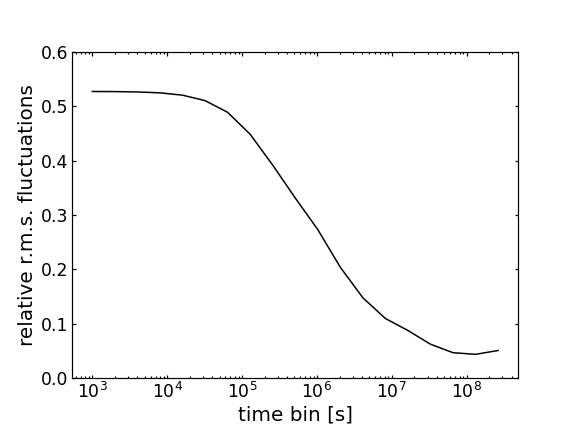}
\caption{The relative root-mean-square of fluctuations for different binning time of NGC $3783$ based on Eq. \ref{equ:psd}.}
\label{fig:lc_amplitude}
\end{figure}

Figure \ref{fig:lc_amplitude} shows that the relative root-mean-square fluctuations decrease significantly as the binning time (hereafter $t_{\rm bin}$) increases from $10^{5}$ to $ 10^{7} \rm \, s$, while the variations on timescale $\rm \leq 10^4 \, s$ remain nearly constant. 
Therefore, we 1) sample the light curve with bin size of $\rm 10^4 \, s$ and length of $\rm 10^{10} \, s$ or approximately $ 10^6 \, t_{\rm bin}$, 2) and rebin the light curve to $10^5$ and/or $10^6 \, \rm s$ bins for simulations with lower gas densities.
For any densities of interest, the sampling time bin $t_{\rm bin}$ is shorter than the typical recombination timescale of the WA.
%depending on a) not much variabilities $\rm \approx 10^4 s$, b)good spectra, c) shorter than the typical $ t_{\rm rec}$ of the WA and duration simulation timescale.

%===========================
% Fig: ngc 3783 lc amplitue
\subsection{10 WA components }\label{subsect:WA 10 componets prototype}

We use the modelling based on the extensive observations with Chandra and XMM-Newton \citep{Mehdipour2017A&A}.
By analyzing the archival Chandra and XMM-Newton grating data taken in $2000$, $2001$ and $2013$ when the AGN was in an unobscured state, \cite{Mao2019A&A} found $9$ photoionized absorption components with different ionization parameters and kinematics. 
We incorporate them as components $1-9$ in our current work.
In the obscured state of the December $2016$ data, \cite{Mehdipour2017A&A} find evidence of a strong high-ionisation component (abbreviation HC in their Fig. 4), which is designated as component $0$ in our present work. 
AGN SEDs do not normally change their shape a lot, unless there is strong X-ray obscuration, like in the above case. 
But the component $0$, due to its high-ionization state, is likely located near the center and therefore is not shielded by the obscurer. 
So it is reasonable to assume that both component $0$ and components $1-9$ experience the same unobscured SED.

%Thereby the 10 WA \textbf{components are} following the same unobscured SED as shown in Fig \ref{fig:sed} and we collect these WA components properties in Table \ref{table:pion good-fit par} (i.e. Table. 1 of \citealp{Mehdipour2017A&A} \& Table. 3 of \citealp{Mao2019A&A}) which can be the initial parameters of TPHO calculation in Sect \ref{subsect:TPHO calc.} as well as PION.
Table \ref{table:pion good-fit par} shows the parameters of the $10$ WA components obtained with the \texttt{pion} model of SPEX. We use them as initial conditions for our \texttt{tpho} calculations.

%%%%%%%%%%%%%%%%%%%%%%%%%%%%%%%%%%%%%%%%%%%%%%%%%%
\begin{table}
\caption{The \texttt{pion} good-fit parameters of $10$ X-ray WAs components of NGC 3783 (see \citealp{Mehdipour2017A&A} Table 1 $\&$ \citealp{Mao2019A&A} Table 3). 
$N_{\rm H}$ is column density in unit of $10^{26} \, \rm m^{-2}$, $\rm log \xi$ is ionized parameter in logarithm scale.
$v_{\rm out}$ is the outflow velocity, $\sigma_{v}$ is RMS velocity.}
\label{table:pion good-fit par}
\begin{tabular}{ p{1cm}|p{1.6cm}|p{1.4cm}|p{1.4cm}|p{1.3cm} }
% \hline
 \hline
 Comp&  $N_{\rm H}$& log$\xi$& $v_{\rm out}$& $\sigma_{v}$ \\
     & [$10^{26} \, \rm m^{-2}$]&    [nWm]&  [$\rm km \, s^{-1}$]&  [$\rm km \, s^{-1}$] \\
 \hline
   0&       16&     3.61&      -2300&     2500 \\ 
   1&       1.11&    3.02&      -480&      120   \\   
   2&       0.21&     2.74&      -1300&     120    \\      
   3&       0.61&     2.55&      -830&      46  \\          
   4&       1.24&    2.40&      -460&      46 \\            
   5&       0.5&     1.65&      -575&      46 \\           
   6&       0.12&     0.92&      -1170&     46 \\          
   7&       0.015&     0.58&      -1070&     46  \\           
   8&       0.007&       -0.01&     -1600&     790 \\          
   9&       0.044&     -0.65&     -1100&     790 \\  
 \hline
\end{tabular}
\end{table}
%%%%%%%%%%%%%%%%%%%%%%%%%%%%%%%%%%%%%%%%%%

\subsection{Recombination timescale}
%We execute \textbf{PION} first to calculate the thermal stability of NGC 3783 source and then to calculate individually where the 10 WA components is in the panel of 
%To provide insight into the initial states of the system, we utilise the PION model to plot a balanced solution of 10 components on a logT - log$\Xi$ thermal stability. 
Photoionization or recombination of the ionized gas in response to changes in the ionising continuum will occur.
For each specific ion $ X_{\rm i}$, the recombination timescale $t_{\rm rec}$ is the time that gas needs to respond to a decrease of the continuum.
It depends on the local density and the ion population, following the equation (\citealp{Bottorff2000ApJ}, \citealp{Ebrero2016A&A}):

\begin{equation}\label{equ:trec}
 t_{\rm rec}(X_{\rm i})=\left(\alpha_r(X_{\rm i}) n_{\rm e} \left[\frac{f(X_{\rm i+1})}{f(X_{\rm i})} - \frac{\alpha_r(X_{\rm i-1})}{\alpha_r(X_{\rm i})} \right] \right)^{-1},
\end{equation}
where $\alpha_r(X_{\rm i})$ is the recombination rate from ion $ X_{\rm i-1}$ to ion $X_{\rm i}$, and $ f(X_{\rm i})$ is the fraction of element $ X $ at the ionization level $i$. 
The recombination rates $ \alpha_r$ are known from atomic physics, and the fractions $f$ can be determined from the ionization balance of the source.
$n_{\rm e}$ is the electron density.

For a cloud with known ionization structure, \(  t_{\rm rec} \propto 1/n_{\rm e}\).
We show in Table \ref{table: ion column density} the column densities of the most relevant ions for each WA component, calculated using the model presented in Table \ref{table:pion good-fit par} in an equilibrium state.
\cite{Ebrero2016A&A} use the ionic column densities of these most important ions to compute $ n_{\rm e} \times t_{\rm rec}$, and they averaged the $ n_{\rm e} \times t_{\rm rec}$ of all used ion to estimate it for the relevant WA component.
The recombination timescale is defined this way through Eq. \ref{equ:trec}, however, changes sign near the ion with the highest relative concentration and becomes very large in the absolute value.
This makes this definition of the recombination timescale not very useful.
%But near the peak concentrations the quantity of differentiating ion concentration with respect to time equals zero, which thus making the time scale very large and unrealistic.

%%%%%%%%%%%%%%%%%%%%%%%%%%%%%%%%%%%%%%%%%%%%%%%%%%%%%
\begin{landscape}
\begin{table}
\caption{
Ion column density $ N_{\rm X}$ in $[10^{20}\, \rm m^{-2}]$ of the $10$ WA components for the specified Fe, S, Si, Mg, O, N ions. 
The results are obtained from our \texttt{pion} calculation and $N_{\rm H}$ and $\rm log \xi$ of Table \ref{table:pion good-fit par} as the input value.
For each component and each element, we show the column densities of the top three maxima abundance ions.}
\label{table: ion column density}
\begin{adjustbox}{max width=25cm}
\begin{tabular}{rrr|rrr|rrr|rrr|rrr|rrr|rrr|rrr|rrr|rrr}
\hline
%\multicolumn{18}{c}{ion column density $\rm N_{X}$ in unit of [$\rm 10^{20} \, \, m^{-2}$]} \\
\hline
 \multicolumn{3}{c}{Comp 0 } &\multicolumn{3}{c}{Comp 1 } &\multicolumn{3}{c}{Comp 2} & \multicolumn{3}{c}{Comp 3} & \multicolumn{3}{c}{Comp 4} & \multicolumn{3}{c}{Comp 5} & \multicolumn{3}{c}{Comp 6} & \multicolumn{3}{c}{Comp 7} & \multicolumn{3}{c}{Comp 8} & \multicolumn{3}{c}{Comp 9}\\
 \hline
 %[$\rm 10^{20} \, \, m^{-2}$] &\multicolumn{2}{c}{[$\rm 10^{20} \, \, m^{-2}$]} &\multicolumn{2}{c}{[$\rm 10^{20} \, \, m^{-2}$]} & \multicolumn{2}{c}{[$\rm 10^{20} \, \, m^{-2}$]} & \multicolumn{2}{c}{[$\rm 10^{20} \, \, m^{-2}$]} & \multicolumn{2}{c}{[$\rm 10^{20} \, \, m^{-2}$]} & \multicolumn{2}{c}{[$\rm 10^{20} \, \, m^{-2}$]} & \multicolumn{2}{c}{[$\rm 10^{20} \, \, m^{-2}$]} & \multicolumn{2}{c}{[$\rm 10^{20} \, \, m^{-2}$]} & \multicolumn{2}{c}{[$\rm 10^{20} \, \, m^{-2}$]}\\
Z& ion& $N_{\rm X} $& Z& ion& $N_{\rm X} $& Z& ion& $N_{\rm X} $& Z& ion& $N_{\rm X} $& Z& ion& $N_{\rm X} $& Z& ion& $N_{\rm X} $& Z& ion& $N_{\rm X} $& Z& ion& $N_{\rm X} $& Z& ion& $N_{\rm X} $& Z& ion& $N_{\rm X} $ \\
 \hline
Fe & 27  & 322.37 & Fe & 25  & 15.45 & Fe & 20  & 1.91 & Fe & 19  & 7.17 & Fe & 19  & 15.50 & Fe & 10  & 5.68 & Fe & 8  & 2.91 & Fe & 8  & 0.40 & Fe & 8  & 0.13 & Fe & 7  & 0.57 \\
Fe & 26  & 167.41 & Fe & 24  & 7.08 & Fe & 21  & 1.76 & Fe & 20  & 5.38 & Fe & 18  & 12.73 & Fe & 9  & 4.63 & Fe & 9  & 0.71 & Fe & 7  & 0.05 & Fe & 7  & 0.08 & Fe & 6  & 0.54 \\
Fe & 25  & 31.98 & Fe & 23  & 4.50 & Fe & 19  & 1.29 & Fe & 18  & 3.42 & Fe & 20  & 5.68 & Fe & 11  & 3.04 & Fe & 10  & 0.15 & Fe & 9  & 0.04 & Fe & 6  & 0.01 & Fe & 8  & 0.19 \\
S & 17  & 252.27 & S & 17  & 10.08 & S & 16  & 1.52 & S & 15  & 5.00 & S & 15  & 9.71 & S & 11  & 2.50 & S & 8  & 0.98 & S & 8  & 0.14 & S & 7  & 0.05 & S & 4  & 0.36 \\
S & 16  & 7.11 & S & 16  & 6.64 & S & 15  & 1.20 & S & 16  & 3.17 & S & 14  & 3.05 & S & 10  & 2.41 & S & 9  & 0.66 & S & 7  & 0.06 & S & 8  & 0.02 & S & 6  & 0.10 \\
S & 15  & 0.10 & S & 15  & 1.25 & S & 17  & 0.58 & S & 14  & 0.78 & S & 16  & 2.95 & S & 9  & 1.44 & S & 7  & 0.17 & S & 9  & 0.03 & S & 6  & 0.02 & S & 5  & 0.09 \\
Si & 15  & 609.58 & Si & 15  & 32.54 & Si & 14  & 3.68 & Si & 14  & 10.88 & Si & 13  & 23.76 & Si & 9  & 6.85 & Si & 8  & 2.57 & Si & 7  & 0.29 & Si & 6  & 0.12 & Si & 5  & 0.79 \\
Si & 14  & 7.14 & Si & 14  & 9.46 & Si & 15  & 3.11 & Si & 13  & 7.49 & Si & 14  & 16.42 & Si & 10  & 6.26 & Si & 7  & 1.45 & Si & 8  & 0.20 & Si & 7  & 0.10 & Si & 6  & 0.59 \\
Si & 13  & 0.05 & Si & 13  & 0.78 & Si & 13  & 1.26 & Si & 15  & 4.63 & Si & 15  & 3.39 & Si & 8  & 2.77 & Si & 9  & 0.41 & Si & 6  & 0.07 & Si & 5  & 0.03 & Si & 4  & 0.14 \\
Mg & 13  & 632.70 & Mg & 13  & 39.29 & Mg & 13  & 5.38 & Mg & 13  & 10.88 & Mg & 12  & 23.91 & Mg & 11  & 6.59 & Mg & 7  & 1.98 & Mg & 7  & 0.24 & Mg & 6  & 0.12 & Mg & 5  & 0.88 \\
Mg & 12  & 2.79 & Mg & 12  & 4.64 & Mg & 12  & 2.62 & Mg & 12  & 10.58 & Mg & 11  & 12.92 & Mg & 9  & 5.05 & Mg & 8  & 1.40 & Mg & 6  & 0.22 & Mg & 5  & 0.12 & Mg & 4  & 0.63 \\
Mg & 11  & 0.01 & Mg & 11  & 0.15 & Mg & 11  & 0.33 & Mg & 11  & 2.71 & Mg & 13  & 11.86 & Mg & 10  & 3.70 & Mg & 6  & 0.72 & Mg & 8  & 0.07 & Mg & 7  & 0.03 & Mg & 6  & 0.19 \\
O & 9 & 9682.56& O & 9  & 665.86 & O & 9  & 122.21 & O & 9  & 341.28 & O & 9  & 638.38 & O & 8  & 153.22 & O & 7  & 47.82 & O & 7  & 4.19 & O & 5  & 1.62 & O & 4  & 17.14 \\
O & 8 & 2.81& O & 8  & 6.04 & O & 8  & 4.87 & O & 8  & 27.52 & O & 8  & 108.38 & O & 7  & 89.37 & O & 6  & 12.09 & O & 6  & 2.67 & O & 4  & 1.37 & O & 3  & 4.61 \\
--& --& --& O & 7  & 0.03 & O & 7  & 0.04 & O & 7  & 0.46 & O & 7  & 3.86 & O & 9  & 57.77 & O & 8  & 7.82 & O & 5  & 1.63 & O & 6  & 0.85 & O & 5  & 4.15 \\
N & 8 & 1306.38& N & 8  & 90.32 & N & 8  & 16.87 & N & 8  & 48.22 & N & 8  & 94.58 & N & 7  & 19.21 & N & 6  & 6.25 & N & 6  & 0.85 & N & 6  & 0.20 & N & 4  & 1.83 \\
N & 7 & 0.14& N & 7  & 0.32 & N & 7  & 0.27 & N & 7  & 1.59 & N & 7  & 6.59 & N & 8  & 17.44 & N & 7  & 2.75 & N & 5  & 0.17 & N & 5  & 0.17 & N & 3  & 1.14 \\
--& --& --& N & 6  & 0.00 & N & 6  & 0.00 & N & 6  & 0.01 & N & 6  & 0.09 & N & 6  & 4.15 & N & 5  & 0.47 & N & 7  & 0.15 & N & 4  & 0.17 & N & 5  & 0.47 \\

\hline
\end{tabular}
\end{adjustbox}
\end{table}
\end{landscape}
%%%%%%%%%%%%%%%%%%%%%%%%%%%%%%%%%%%%%%%%%%%%%%%%%%%%%%%%%%%%%%%%%%%

%%%%%%%%%%%%%%%%%%%%%%%%%%%%%%%%%%%%%%%%%%%%%%%%%%
\begin{table}
\caption{
Ion species recombination timescale $ t_{\rm rec}$ and element charge time $t_{ C}$ using \texttt{pion} calculation for WA component $0$ with $\rm log \xi = 3.61$ as well as $n_{\rm H} = 10^{10} \, \rm m^{-3}$. }
\label{table:t_C and t_rec.}
\begin{tabular}{ p{2cm}|p{2cm}|p{2cm}}
 \hline
 \multicolumn{3}{c}{WA component 0 of NGC 3783} \\
 \hline 
 \hline 
 ion &   $t_{\rm rec} \, [\rm s]$&  $ t_{ C} \, [\rm s]$ \\
 %    &           [s]&     [s]  \\
 \hline
 \ion{Fe}{XXVII}&   $-2.23 \times 10^{7}$&   \\
 \ion{Fe}{XXVI}&    $1.98 \times 10^{7}$&      \\
 \ion{Fe}{XXV}&    $6.17 \times 10^{6}$&   $2.45 \times 10^{7}$  \\
 \hline
 \ion{S}{XVII}&    $-8.13 \times 10^{7}$&      \\
 \ion{S}{XVI}&    $2.38 \times 10^{7}$&   $8.07 \times 10^{7}$   \\
 \hline
 \ion{Si}{XV}&    $-1.18 \times 10^{8}$&      \\
 \ion{Si}{XIV}&    $1.40 \times 10^{6}$&   $1.17 \times 10^{8}$   \\
 \hline
 \ion{Mg}{XIII}&    $-1.82 \times 10^{8}$&      \\
 \ion{Mg}{XII}&    $8.10 \times 10^{5}$&   $1.81 \times 10^{8}$   \\
 \hline
 \ion{O}{IX}&    $-6.00 \times 10^{8}$&   $5.99 \times 10^{8}$   \\
 \hline
\end{tabular}
\end{table}
%%%%%%%%%%%%%%%%%%%%%%%%%%%%%%%%%%%%%%%%%%

Hence we simply define the recombination time scale of an element by the time that the ionized plasma needs to change the average charge of the element by one charge state, namely:
\begin{equation}\label{equ:trec from ave-char}
     t_{C} = \frac{1}{dC/dt} \, \,  .
\end{equation}
The average charge $ C$ is defined by
\begin{equation}\label{equ:ave-char}
     C = \frac{\sum_{i=1}^{n} f(X_i) \times (i-1)}{\sum_{i=1}^{n} f(X_{i})} \, \, ,
\end{equation}
and
\begin{equation}\label{equ:ave-char/t}
     \frac{dC}{dt} = \sum_{i=1}^{n} [f(X_{\rm i+1}) \alpha_r (X_{\rm i+1}) - f(X_{\rm i}) \alpha_r (X_{\rm i})] i \, \, ,
\end{equation}
where the denotation of $f(X_{\rm i})$ and $\alpha_{r}(X_{i})$ are same as shown in the Eq. \ref{equ:trec}.
%Here the time scale $ t_{C}$ means that ionized plasma is needed to change the average charge of the element by one charge state.
Table \ref{table:t_C and t_rec.} shows examples of $ t_{\rm C}$ and $ t_{\rm rec}$ where consistency between the two values can be seen. 
Following the above calculation, we replace the previous definition of $t_{\rm rec}$ by $t_{\rm C}$ to characterize the recombination timescale. 

To quantity the relation between $ t_{\rm C}$ and $ n_{\rm e}$, we use the logarithm of their product using the following equation
\begin{equation}
      \rm log(\it t_{C} \times n_{\rm e}) = K(\xi, Z) \, \, ,
\end{equation}
where we express $t_{\rm C}$ in second and $n_{\rm e}$ in $\rm m^{-3}$.
%\propto
%Where the $K(\xi, Z)$ is different constants corresponding to different $\rm log \xi$ and ion population.
%It means for each WA component with one initial $\rm log \xi$, the $\rm C_{log \xi, ion}$ dependents on which ion we choose. 
Table \ref{table:the constant value} shows the $K(\xi, Z)$ value for the 10 WA components from a \texttt{pion} calculation using Eq. \ref{equ:trec from ave-char} for Fe, S, Si, Mg, O, N.

%%%%%%%%%%%%%%%%%%%%%%%%%%%%%%%%%%%%%%%%%%%%%%%%%%
\begin{table}
\caption{
The quantity $K(\xi, Z)$ for the 10 WA components.
}
\label{table:the constant value}
\begin{tabular}{ p{0.7cm}|p{0.95cm}|p{0.88cm}|p{0.95cm}|p{1.1cm}|p{0.92cm}|p{0.90cm} }
 \hline
 Comp & $K(\xi, \rm Fe)$&  $K(\xi, \rm S)$& $K(\xi, \rm Si)$&  $K(\xi, \rm Mg)$& $K(\xi, \rm O)$& $K(\xi, \rm N)$ \\
 \hline
 0&  17.39&  17.91&  18.23&  18.26&  18.78&  18.95 \\
 1&  16.58&  17.22&  17.33&  17.47&  17.91&  18.06 \\
 2&  16.10&  16.96&  17.25&  17.18&  17.57&  17.71 \\
 3&  16.13&  16.80&  16.99&  17.11&  17.46&  17.59 \\
 4&  16.13&  16.40&  16.77&  16.98&  17.30&  17.42 \\
 5&  15.31&  16.06&  15.98&  16.23&  17.00&  17.08 \\
 6&  15.74&  16.29&  16.22&  16.22&  16.71&  16.94 \\
 7&  16.02&  16.47&  16.42&  16.30&  16.60&  16.84 \\
 8&  16.43&  16.56&  16.71&  16.52&  16.55&  16.68 \\
 9&  16.70&  16.63&  16.94&  16.84&  16.76&  16.73 \\
 \hline
\end{tabular}
\end{table}
%%%%%%%%%%%%%%%%%%%%%%%%%%%%%%%%%%%%%%%%%%

\subsection{TPHO calculation}\label{subsect:tpho calc.}
\begin{figure}[!tbp]
\centering
\includegraphics[width=1.05\linewidth]{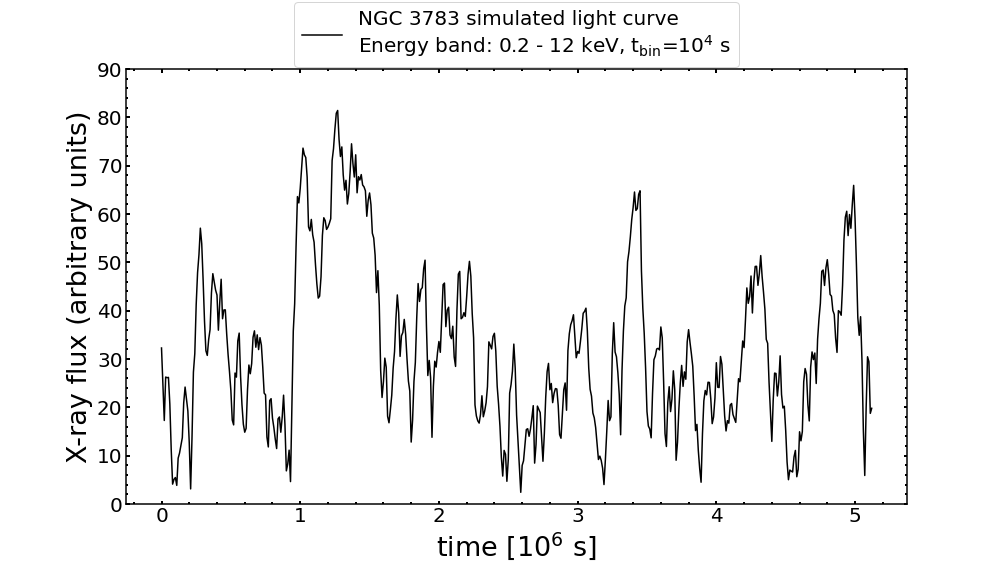}
\caption{NGC 3783 simulated light curve in the $0.2 - 12$ keV band from the PSD shape of \cite{Markowitz2005ApJ}.}
\label{fig:lc_E4}
\end{figure}

We now apply the \texttt{tpho} model to calculate the time evolution of each WA component of NGC 3783.
Starting point of the \texttt{tpho} calculation is an equilibrium solution which can be obtained from the \texttt{pion} model, ideally using the luminosity averaged over the entire simulation.
Figure \ref{fig:lc_E4} shows the simulated light curve with a $t_{\rm bin}$ of $10^{4}$ s, $512$ data points resulting in a duration of $5.12 \times 10^{6} \rm \, s$.
The initial parameters, such as ionization parameter $\xi$ and column density $ N_{\rm H}$, are taken from Table \ref{table:pion good-fit par}. 
%Here we collect these useful parameters value to Table\ref{table:pion good-fit par} adding the outflow velocity.
We explore a range of hydrogen number density ($ n_{\rm H}$) from $10^8 \, \rm m^{-3}$ to $10^{15} \, \rm m^{-3}$.

%============================
%  Fig: ngc 3783 lc
%
%============================

If the photoionized plasma is too far away from the nucleus and/or the density of the plasma is too low, the plasma is in a quasi-steady state with its ionization state varying slightly around the mean value corresponding to the mean ionizing flux level over time; whereas for high density the ion concentration will simultaneously follow the ionizing luminosity variation in a near-equilibrium state (\citealp{Krolik&Kriss1995ApJ}; \citealp{Nicastro1999ApJ}; \citealp{Kaastra2012A&A}; \citealp{Silva2016A&A}).
Here we select the density range from $10{^8} \, \rm m^{-3}$ to $10^{15} \, \rm  m^{-3}$ that encompasses the steady state, delayed state and nearby equilibrium state (see Sect. \ref{subsect:delayed state}) as follows from our \texttt{tpho} calculation.

To further investigate the delayed state and to apply density diagnostics, it is needed to set up the simulation with the following conditions,
\begin{equation}\label{equ:tbin empirial relationship}
 t_{\rm bin} << t_{\rm var} (\sim t_{\rm C}) << t_{\rm tot} \, \, .  
\end{equation}
It is necessary to have $t_{\rm bin} << t_{\rm var}$, which guarantees that the variability of the ionizing luminosity is sufficiently sampled.
With $t_{\rm var} << t_{\rm tot}$, each simulation run would then contain multiple ionization-recombination periods to get better constraints.
As will be addressed below, significant lags between the outflows and the central source are likely to be observed when $t_{\rm var}$ approaches the recombination time $t_{\rm C}$.

To be specific, we use $t_{\rm bin} = 10^4, 10^5, 10^6 \, \rm s$ for density $10^{12}, 10^{11}, 10^{10} \, \rm m^{-3}$, respectively.

%----------------------------------------------------------------------
\section{Results}\label{sect:3}
\begin{figure}[htbp]
\centering
\includegraphics[width=\linewidth]{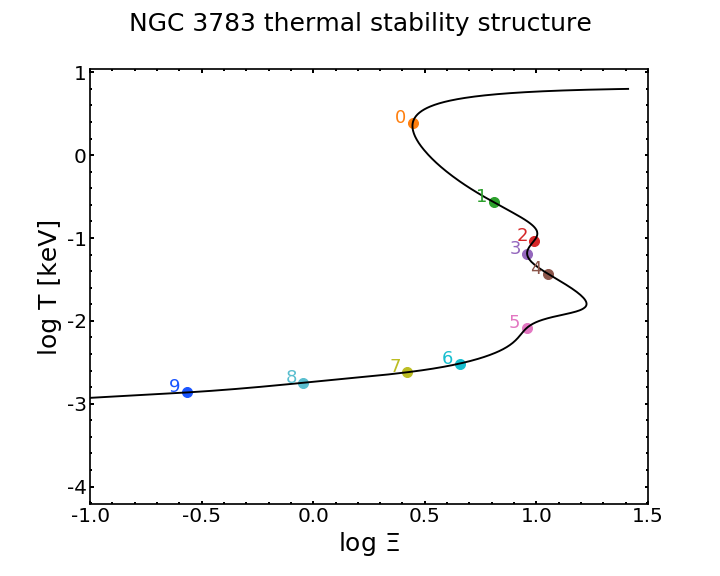}
\caption{
The electron temperature in $\rm log \it T$ versus pressure ionization parameter in $\rm log \Xi$ ("S" curve in black).
The colored dots show the $10$ X-ray WA components.
%Component 2 and component 4 are in unstable equilibrium state.
}
\label{fig:scurve}
\end{figure}

We calculate the thermal stability curve of the $10$ WA components by computing the temperature for a range of $\xi$ values using the \texttt{pion} model made with our SED.
%Then we change log$\xi$ initial value from -7 to 5.0 and make other parameters in default as the \emph{pion} have in order to get a array of logT-log$\Xi$ panel (i.e. "S" curve) as Figure \ref{fig:scurve} black curve shown.
%In Figure \ref{fig:scurve}, $\rm T_e$ is the electron temperature, and $\Xi$ represents 
From this, we determine the pressure ionization parameter, the ratio of radiation pressure to gas pressure, 
$\Xi$ defined as (\citealp{Krolik1981ApJ})
\begin{equation}\label{equ:big xi}
    \Xi = L/4\pi r^2 c p = \xi / 4 \pi c k T \, \, .
\end{equation}
The shape of this curve in general depends on the precise form of the incoming X-ray spectrum.
In this equation, $p$ is the gas pressure, $c$ is the speed of light, $k$ is the constant of Boltzmann, and $T$ is the electron temperature.
Figure \ref{fig:scurve} shows a plot of $\rm log \it T$ versus $\rm log \Xi$.
Note that in general for a single value of $\Xi$ more temperature solutions are possible (in particular when log$\Xi$ is close to $1$).
Only solutions with $dT / d\Xi > 0$ are stable.
Hence it seems that components $1$ and $4$ are in an unstable state.

%============================
%  Fig: density vs. xil
%
%============================

Using the light curve described in Sect. \ref{subsect:lc} and the SED described in Sect. \ref{subsect:sed}, we apply the \texttt{tpho} model for a range of densities for each WA component.
The ion concentration is changing over time because the gas ionises when the flux rises, and recombines when the flux goes down. 
The recombination is responsible for the lag between the light curve and the ion concentration variability. 
We also investigate the relationship between density and lag time scales to further characterize this phenomenon.
Here we present the Component 0 results in more detail as an example.

\subsection{Delayed state in ion concentration}\label{subsect:delayed state}

Due to its high ionization state, Component $0$ predominately produces absorption lines in the Fe-K band, hence we focus on Fe.
We use the light curve with $ t_{\rm bin}=10^4 \, \rm s$ for the density ranging from $10^8 \, \rm m^{-3}$ to $10^{15} \, \rm m^{-3}$ in this section.

%In Fig. \ref{fig:Fe_last6_3nH}, for the top two rows panels, we plot the \ion{Fe}{XXVII} to \ion{Fe}{XXIV} concentration as a function of time.  
%The \ion{Fe}{XXVII} ion concentration shows clear variability at $1 \times 10^{12} \rm m^{-3}$ which seems to follow the evolution of the light curve with a lag.
%This lag is more visible when the ionising flux decreases.
%For a relatively low density of $1 \times 10^{10} \rm m^{-3} $, the \ion{Fe}{XXVII} ion concentration exhibits almost no variability.
%And for a higher density of $1 \times 10^{14} \rm m^{-3}$ , the variability of \ion{Fe}{XXVII} become well synchronized with the light curve as shown in this Figure.
%The \ion{Fe}{XXVI} - \ion{Fe}{XXIV}, ion concentrations exhibit inverse pattern and lagged state.
%Following Eq. \ref{equ:trec}, the \ion{Fe}{XXVI} concentration is affected both by recombining into \ion{Fe}{XXV}, and ionizing into \ion{Fe}{XXVII}.
%The last panel of Fig. \ref{fig:Fe_last6_3nH} shows the time-evolved electron temperature $T$ profile for three different densities same as above. 
%Both time-evolved ion concentration and temperature show the same behaviour going through low, moderate and high densisy, which are also in agreement with the calculation of \cite{Rogantini2022ApJ} shown.

\begin{figure*}
\centering
\includegraphics[width=0.9\textwidth]{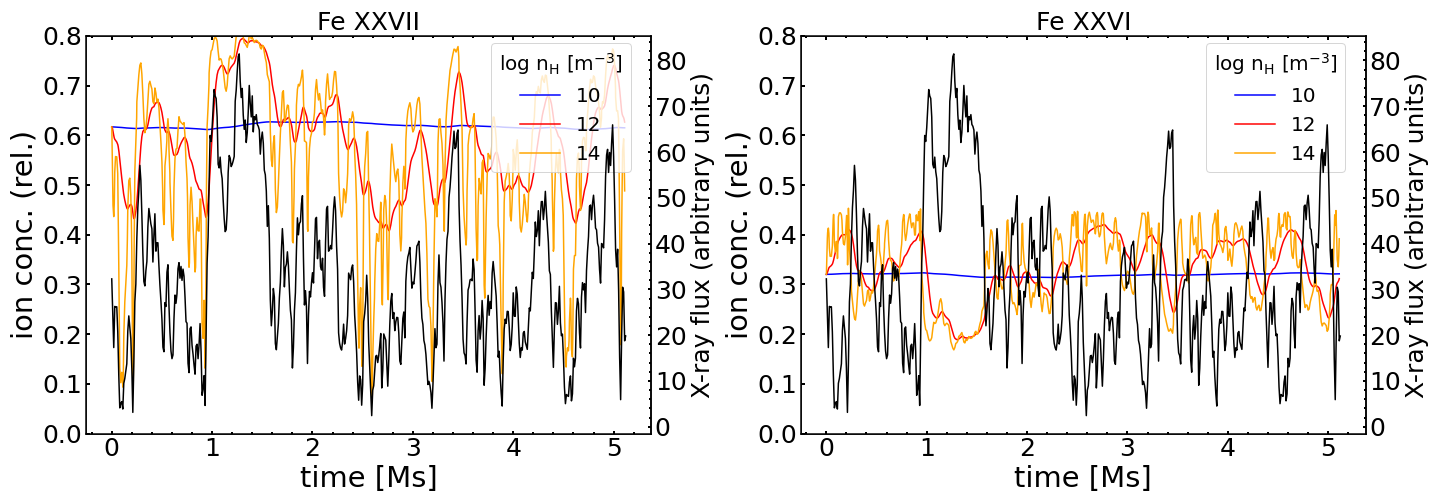}
\includegraphics[width=0.9\textwidth]{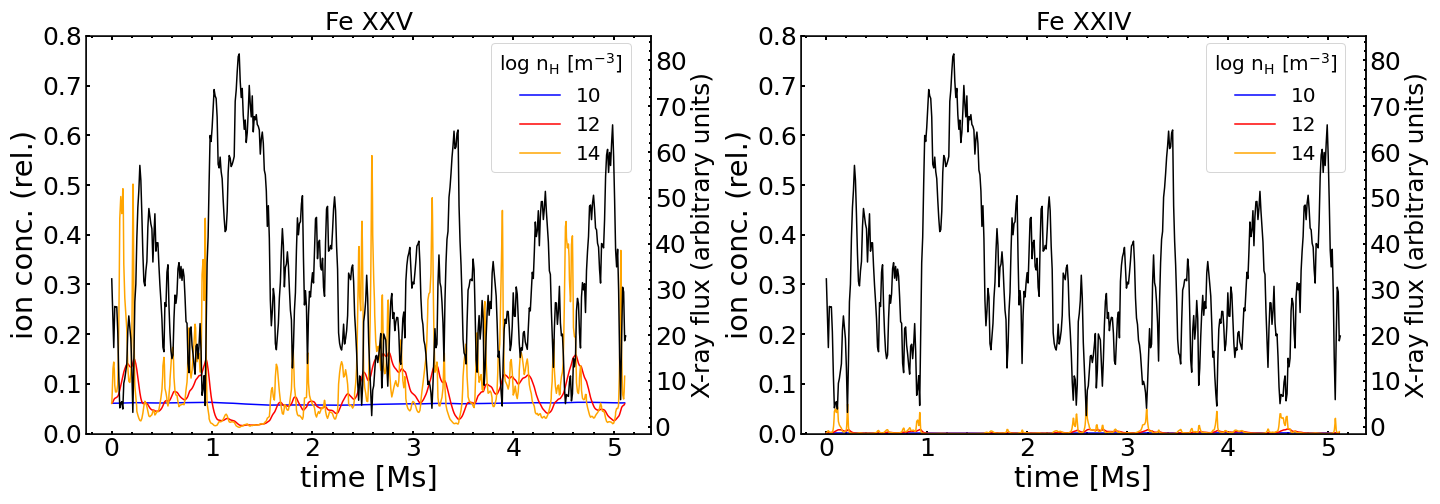}
\includegraphics[width=9cm]{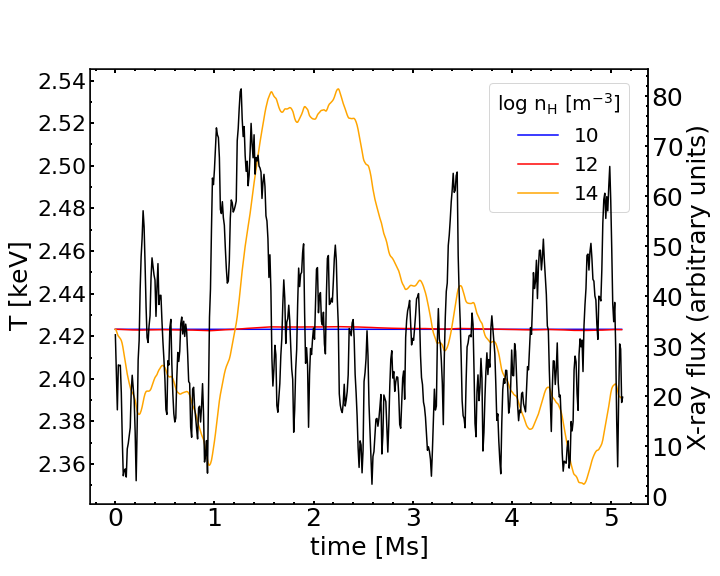}
\caption{
The \texttt{tpho} calculation of the X-ray WA component $0$ of NGC $3783$ based on the simulated light curve (Fig. \ref{fig:lc_E4}) and SED (Fig. \ref{fig:sed}).
%The TPHO calculation of the X-ray WA component 0 of NGC 3783 using the $\rm  10^{4} \, s$ time bin size simulated light curve (black line).
The time-dependent ion concentration of \ion{Fe}{XXVII}, \ion{Fe}{XXVI}, \ion{Fe}{XXV}, \ion{Fe}{XXIV} are shown for hydrogen number density of $\rm 10^{10}, 10^{12}, 10^{14} \, m^{-3}$ in blue, red and orange, respectively.
The last figure shows the electron temperature $T$ of the plasma of Component $0$ as a function of time. 
}
\label{fig:Fe_last6_3nH}
\end{figure*}

Fig \ref{fig:Fe_last6_3nH} shows the ion concentrations of the most importannt Fe ions and the temperature as a function of time for three different densities. When the luminosity increases, the \ion{Fe}{XXVII} concentration increases, of the expense of \ion{Fe}{XXIV} - \ion{Fe}{XXVI} due to conservation of the total number of Fe nuclei.

For a density of $1 \times 10^{12} \, \rm m^{-3}$, the \ion{Fe}{XXVII} ion concentration shows clear lag with respect to the light curve.
The lag arises mainly from the recombination timescale of the plasma, which depends inversely on the density (see Eq. \ref{equ:trec}).
Therefore with increasing density, the \ion{Fe}{XXVII} ion concentration (from blue, red to orange curve) will correspond to smaller lag timescale. For a higher density of $1 \times 10^{14} \rm \, m^{-3}$ , the variability of \ion{Fe}{XXVII} becomes well synchronized with the light curve.
That is also the reason why the ion concentrations show smoother appearance from high to low density.
And the standard deviation amplitude of \ion{Fe}{XXVII} for $10^{10}, 10^{12}, 10^{14} \, \rm m^{-3}$, is $0.007$, $0.14$ and $0.27$ respectively.

%\textbf{The smoother of \ion{Fe}{XXVII} ion concentration in lower density (red) than the higher (orange) results from the decrease of amplitude of ionizing flux, which following the relation of \( \Delta F_{\rm ion} = \frac{1}{4 \pi} n_{e} \xi\) in TPHO simulation.So for the 
%}

In Fig. \ref{fig:Fe_last6_3nH}, the last panel shows the time-evolved electron temperature $T$ profile.
For low density, e.g. $10^{10}, 10^{12} \, \rm m^{-3}$, there are no variations and the temperature is constant over time, because cooling rate almost equals the heating rate within the duration of the light curve.
For higher density, e.g. $10^{14} \, \rm m^{-3}$, the electron temperature lags due to the cooling time $\sim 10^6 \, \rm s$.
Both time dependencies of the ion concentration and electron temperature are in agreement with \cite{Rogantini2022ApJ} (see their Figure $7 \, \& \, 8$).

%============================
%  Fig: Fe ionCon 3nH
%
%============================

\subsection{ Density-dependent lag}\label{subsect:density-dependent lag}
\begin{figure}[!tbp]
\centering
\includegraphics[width=\linewidth]{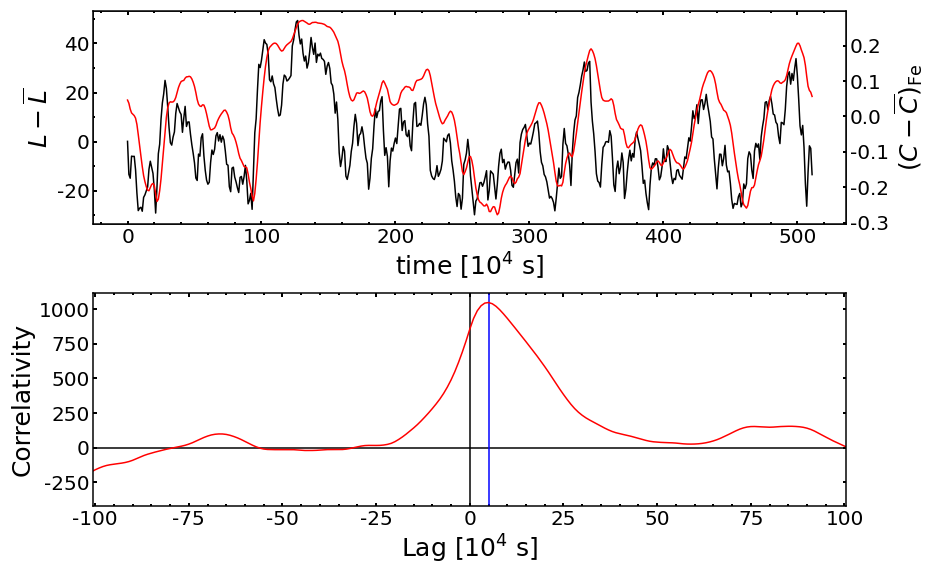}
\caption{
Cross correlation of the Fe average charge with ionizing luminosity for Component 0 of NGC $3783$.
Top panel: Fe mean-subtracted average change (red curve) from \texttt{tpho} calculation with hydrogen number density of $\rm 10^{12} \, m^{-3}$, and with ionizing luminosity in black ($ t_{\rm bin} = 10^{4} \, \rm s$).
Bottom: cross correlation versus lag. 
The blue vertical line marks the lag timescale of $5  \times 10^{4} \, \rm s$ corresponding to the largest correlation.
}
\label{fig:ccf_char-lumi}
\end{figure}

The lag between the ionizing luminosity and ion concentration as shown in Fig. \ref{fig:Fe_last6_3nH} is caused by delayed recombination.
To quantify the lag in our current work, we determine the cross-correlation function (CCF) between the average charge of Fe and the ionizing luminosity as shown in Fig. \ref{fig:ccf_char-lumi}. 
The Fe average charge $C_{\rm Fe}$ is derived using Eq. \ref{equ:ave-char} for each time bin. 
The ionizing luminosity $L$ is also the value for each time bin.
For both quantities, we calculate the relative variation by subtracting the long-term average.
The CCF is a measure of the similarity of two series as a function of the displacement of one relative to the other as shown in the following equation: 
\begin{equation}\label{equa:CCF equation}
    ((L-\overline{L})*(C-\overline{C}))[\Delta t] = \sum_{-\infty}^{\infty} \overline{(L-\overline{L})[t]} (C-\overline{C})[t+\Delta t] \, .
\end{equation}
%Where $f$ and $g$ are discrete functions which can be replaced by luminosity and ion concentration, 
$\Delta t$ is defined as the displacement, also known as the lag.
We compute the CCF using the procedure of the website of Scipy-correlation \footnote[1]{https://docs.scipy.org/doc/scipy/reference/generated/scipy.signal.corr
elation$\_$lags.html}
%\href{https://docs.scipy.org/doc/scipy/reference/generated/scipy.signal.correlation_lags.html}{Scipy-correlation} 
between the quantity \( (C_{\rm j} - \overline{C})_{\rm Fe} \) and mean-subtracted ionizing luminosity (\( L_{\rm j} - \overline{L}\)).
$\overline{C}$ and $\overline{L}$ here are the average value calculated from all bins.
We take the lag value corresponding to the largest correlativity for each calculation (see red curve at the bottom of Fig. \ref{fig:ccf_char-lumi}).

Figure \ref{fig:ccf_char-lumi} shows the CCF for a hydrogen number density of $10^{12} \, \rm m^{-3}$ and component 0. 
We see here a lag timescale of $5 \times 10^{4} \, \rm s$ (16 hours or more than half a day) for a density of $\rm 10^{12} \, m^{-3}$.
There are lag timescale of $1 \times 10^{6}, 2 \times 10^{5} \, \rm s$ for the density of $10^{10}, 10^{11} \, \rm m^{-3}$, respectively. 
The uncertainties in the simulated light curve, derived from the power spectral density of \cite{Markowitz2005ApJ}, can introduce errors in the lag measurement. 
To account for this, we perform re-simulations of the light curves by considering the errors in the observed PSD. 
By comparing the resulting lag times obtained from different light curves, we determine an uncertainty of roughly $10\%$.

%============================
%  Fig: luminosity vs. average charge(t) 
%
%============================
%

\begin{figure}[!h]
\centering
\includegraphics[width=\linewidth]{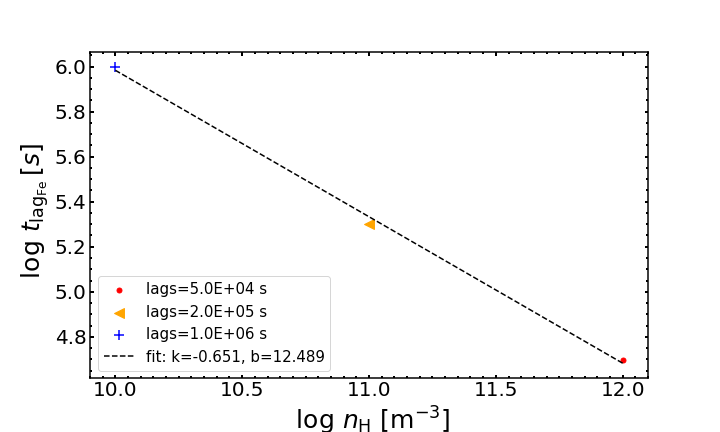}
\includegraphics[width=\linewidth]{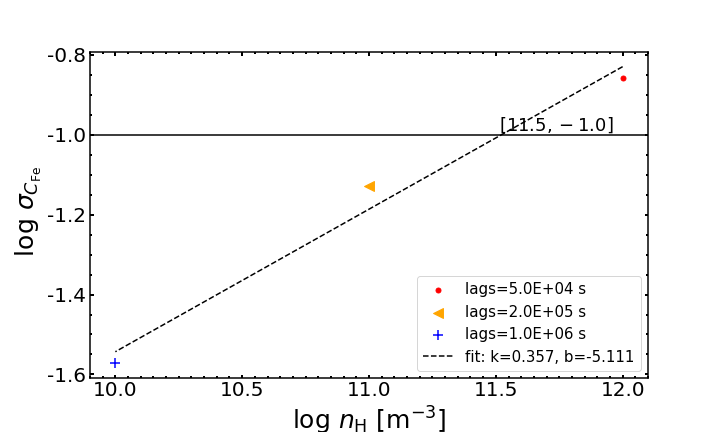}
\caption{
Top: The relation of lag timescale vs. hydrogen number density based on Fig. \ref{fig:ccf_char-lumi} for Component $0$.
Bottom: The relation of $\sigma_{C}$ for Fe vs. density for Component $0$.
}
\label{fig:lags-lognH-sigma}
\end{figure}

The lag timescale in our work detected by the CCF method is the net effect of the recombination and ionization periods.
Similar to $t_{\rm rec} \sim 1/n$, the lag timescale decreases with increasing hydrogen number density due to the shorter recombination timescale as shown in the top panel of Figure \ref{fig:lags-lognH-sigma}.
We fit these three data points (see black dot line) with a power law 
\begin{equation}\label{equ: lag vs. nH}
   t_{lag_{\rm Fe}} = 10^{b} \times n_{\rm H}^{k} \, ,
\end{equation}
and find $\rm k = -0.651$ and $\rm b = 12.489$.
Here, we find for Component 0, $t_{lag_{\rm Fe}} \varpropto 1/n_{\rm H}^{0.651}$. 
The observed lag, which is a result of both recombination and ionization, is described by a power index of less than $1$.
We take here as a typical minimum useful lag the time scale of $\rm 10^4 \, s$, which is approximately the timescale necessary for relevant AGN observations, i.e. the time needed to obtain a spectrum with sufficient quality. 
%
%============================
%  Fig:  
%
%============================
%

To scale the amplitude to the mean-substracted average charge, we use root-mean-square \( \rm \sigma_C = \sqrt{\frac{ \sum_{j=1}^{j=n} (C_{\rm j} - \overline{C})^2}{n}} \), where
$ C_{\rm j}$ means average charge in the $ j^{\rm th}$ time binsize and $ n$ means the numbers of time bins.
Hence, for each calculation from \texttt{tpho}, we further calculate the $\rm \sigma_C$ of the Fe ion concentration.
At the bottom of Figure \ref{fig:lags-lognH-sigma}, we show the relation between $\sigma_{C}$ of Fe and hydrogen number density.
We find that relatively high density corresponds to a relative large $\sigma_C$ value, as well as a shorter lag time.
We fit these three data points (see black dot line) with a power-law 
\begin{equation}\label{equ:sigma vs. nH}
    \sigma_{C_{\rm Fe}} = 10^{\rm b} \times n_{\rm H}^{\rm k} \, ,
\end{equation}
and find $\rm k=0.357$ and $\rm b=-5.111$. 
We assume that we can observe in practice significant change in absorption line spectra where $\sigma_{C_{\rm Fe}} > 0.1$; as for lower densities or longer lags, the amplitude of the charge variations become practically too small for effective measurement.
$10\%$ serves as an empirical standard. A $10\%$ change in average change for one element corresponds typically to changes of order $10\%$ in the ion concentration and hence ionic column densities and line optical depth. This is approximately true based on the observed Fe ion concentrations reported in \cite{Kaspi2002ApJ}.
For component $0$, for Fe, we find a density of (about) $\rm 3 \times 10^{11} \,  m^{-3}$ (see the black horizontal line) when $\sigma = 0.1$.

From the constraints of lag larger than $10^4 \, \rm s$ and $\sigma_{\rm C}$ larger than $10 \%$, for component $0$ and Fe, lags can be detected for densities roughly between $3 \times 10^{11}$ and $\rm 1 \times 10^{13} \, m^{-3}$.

Hence, the above method, dubbed here, \texttt{tpho}-delay method, can be used to constrain the density range of the outflowing wind when the source variability timescale is comparable to the recombination time, i.e. ionized plasma in delayed state.
The same calculation has been carried out for the other $9$ components.

\subsection{Constraint on the WA density}\label{subsect:constraint on WA density}

Apart from the constraint that can be obtained from the spectral-timing analysis, i.e. \texttt{tpho}-delay method of Section \ref{subsect:density-dependent lag} in our current work, we take into account other astrophysical conditions.
A lower limit to the density is obtained by assuming that the thickness ($\Delta r$) of the WA plasma can not exceed its distance ($r$) to the SMBH (\citealp{Krolik2001ApJ}; \citealp{Blustin2005A&A}), namely \( \Delta r \leq r \),
so we can derive the lower limit of the physical density by:
\begin{equation}
    n_{\rm H, low} = \frac{\xi \times N_{\rm H}^{2}}{L} \, ,
\end{equation}
where $N_{\rm H}$ is the column density.

The upper limit to the density can be obtained based on the assumption that the outflow velocity $v_{\rm out}$ of the wind is larger than or equal to the escape velocity ($  v_{\rm esc} = \sqrt{ 2 G M_{\rm BH} / r}$ ), e.g. \cite{Blustin2005A&A}. 
It gives,
%\[ {v_{esc}^2} = \frac{G \times M_{BH}}{r} \]
\begin{equation}
     n_{\rm H, upp} = \frac{L \times {v_{\rm out}^4}}{{4 \times (G \times M_{\rm BH})^2} \times \xi} \, ,
\end{equation}
Using $M_{\rm BH} = 3 \times 10^7 \, M_{\sun}$ (\citealp{Vestergaard2006ApJ}) and other parameters from Table \ref{table:pion good-fit par}, we compute $n_{\rm H, low}$ and $n_{\rm H, upp}$.

\begin{figure*}[!tbp]
\centering
\includegraphics[width=\textwidth]{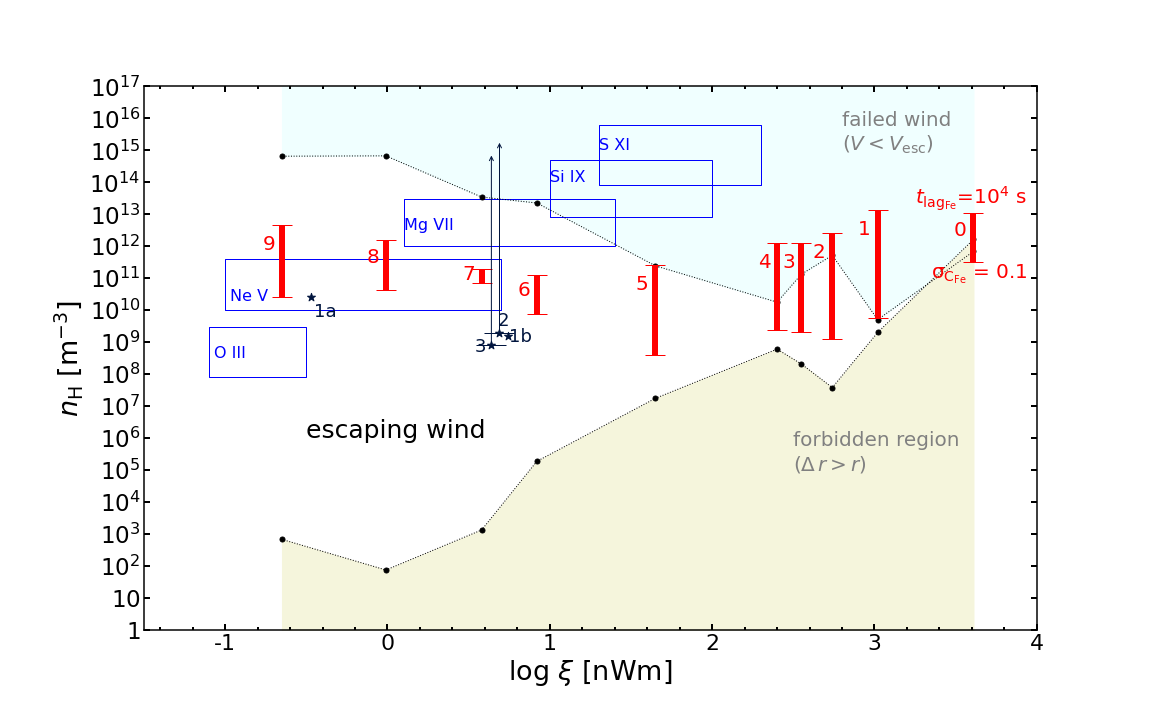}
\caption{
The detectable density ranges of $10$ X-ray WA components in the unobscured state of NGC $3783$. 
The $10$ WAs are marked in red.
Upper black points and line: outflow velocity equals the escape velocity.
For higher densities, winds cannot escape i.e. they are failed winds as shown in the light azure shaded region.
Lower black points and line: the thickness of the wind equals its distance to the core.
For lower densities, no solution exists, i.e. forbidden region shown in the yellow region.
Red vertical solid lines: the detectable density range by the \texttt{tpho}-delay method for all $10$ X-ray WAs.
The upper limit: lag equals $10^4 \rm \, s$.
For higher densities, lags are shorter and more difficult to measure due to photon statistics.
The lower limit: relative variations of the charge of Fe ions is $10 \% $.
For lower densities, variations become too small to measure.
Navy stars and lines with labels $1a$ to $3$: density measurements or lower limits from \cite{Gabel2005ApJ}.
$\xi$ values converted for models with the SED of \cite{Mehdipour2017A&A}.
Rectangular boxes: regions where density-sensitive X-ray lines can be used to measure densities.
}
\label{fig:summary methods}
\end{figure*}

Following the technique introduced in \cite{Mao2017A&A}, we calculate the ranges of detectable densities using absorption lines from meta-stable levels in C- like ions for NGC $3783$.
The parameter range covered by this method exhibits dependence on ionization parameter and density, as depicted in the colored rectangles of Figure \ref{fig:summary methods}.

%============================
%  Fig: density vs. xil
%
%============================

Figure \ref{fig:summary methods} shows a comprehensive constraint on the density ranges for all 10 WA components as derived from the methods described above.
%The numbers in red mark the 10 X-ray components.
%In the TPHO-delay method calculation, the upper limit of the density range is determined by the minimal lag of $\rm 10^4 s$, while the lower limit is set by the minimal amplitude of Fe charge variation $\sim 0.1$.
%The red vertical solid lines are the detectable density ranges for all 10 X-ray WAs where ionized plasma is in a delayed state.

The margins for a failed wind and the forbidden region obtained based on previous observations present the upper and lower limits density for $ V = V_{\rm esc}$ and \(\Delta r = r \), respectively.
For X-ray component $0$, we find that $ 10^{12.2} < n_{\rm H} < 10^{11.8} \, \rm m^{-3}$ i.e. the lower limit is higher than the upper limit.
It means that perhaps X-ray component $0$ is a candidate for a failed wind.

We see that for the components with a relatively narrow physical density range as well as higher ionization parameters, such as components $1 - 4$, the \texttt{tpho}-delay method provides powerful diagnostics for the failed wind scenario. 
The detectable density range of the \texttt{tpho}-delay method for these components $2 - 5$ covers most of their available physical density range, hence their density can be determined well using our method.
Starting from component $5$, metastable density works, \ion{Si}{IX} and \ion{S}{XI} can diagnose whether component $5$ is failed wind or not.

For components $6-9$ with a relatively broad allowed physical density range and a low ionization parameter,
we can combine the \texttt{tpho}-delay method with metastable C-like ions to give density constraints in the X-ray band.
For components $7-9$, we can have a cross-check between \texttt{tpho}-delay and the \ion{Ne}{V} metastable lines. 
\ion{Mg}{VII} and \ion{Ne}{V} as well as \ion{O}{III} metastable lines also expand the detectable density range for components $6-9$.

Gray stars and lines represent the density measurements or lower limit from UV observations using the UV metastable absorption lines (Table $3$ in \citealp{Gabel2005ApJ}). 
Their ionization parameter $U_{\rm H} = Q_{\rm H} / 4 \pi c n_{\rm H} r^2$ was converted into $\xi$ for the models with the SED and luminosity of \cite{Mehdipour2017A&A}. 
%We adopt $n_{\rm H} = 1.2 n_e$ considering solar abundance distribution and fully ionized.
We will discuss UV components further in Section \ref{sect:5}.

We note that here we use $t_{lag_{\rm Fe}}$, the lag timescale of the Fe average charge, to give the detectable density range for components at different ionization states.
However, the other elements would complement Fe in particular for the low ionization components.
To fully comprehend the behaviour of the WA outflows, we will need to conduct similar calculations for other critical elements, such as C, O, N, Mg and Si, among others.

\subsection{TPHO vs. PION}\label{subsect:TPHO vs. pion}

%============================
%  Fig: density vs. xil
\begin{figure*}[!tbp]
\centering
\includegraphics[width=0.7\textwidth]{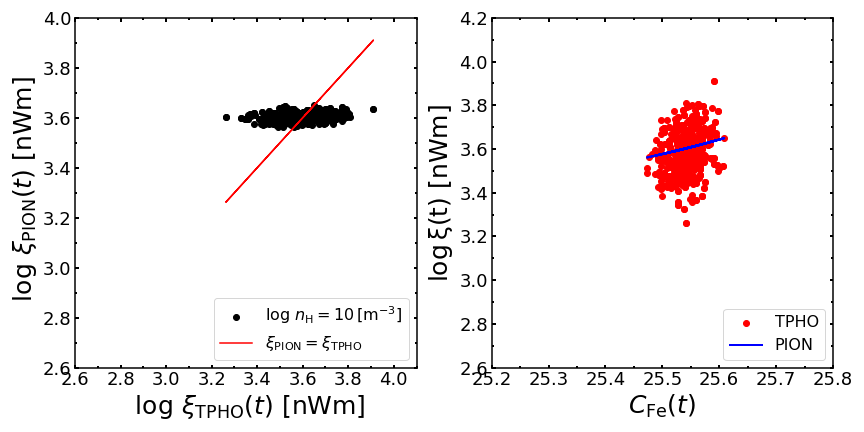}
\includegraphics[width=0.7\textwidth]{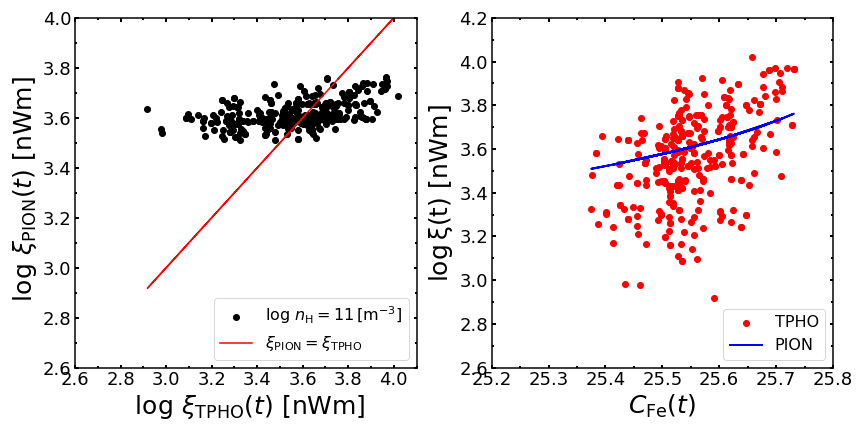}
\includegraphics[width=0.7\textwidth]{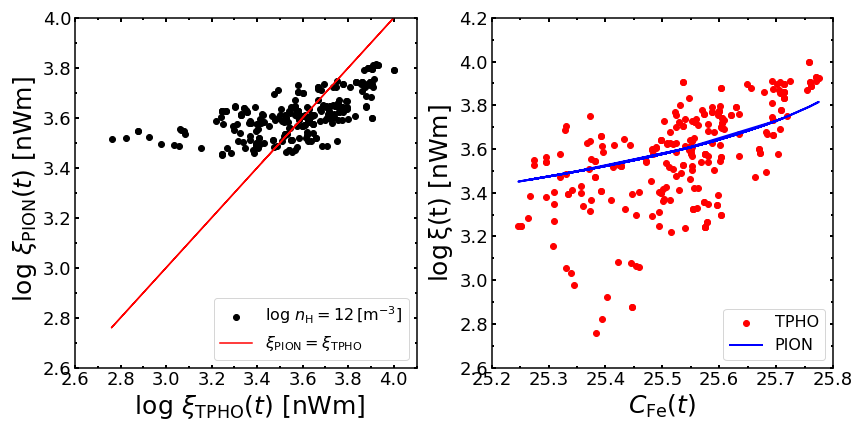}
\caption{
Ionization parameter and Fe ion average charge comparisons for \texttt{tpho} and \texttt{pion} for X-ray WA component $0$ of NGC $3783$ for three different densities ($10^{10}, 10^{11}$ and $ 10^{12} \, \rm m^{-3}$ from top to bottom).
In the left panels: the black dots represent the $\xi$ values of \texttt{pion} that are needed to obtain are the same Fe average charge as for the \texttt{tpho} model. 
$\xi_{\rm TPHO}$ is a luminosity-dependent quantity in \texttt{tpho} calculations (See Eq. \ref{equ: TPHO xi}).
The red line represents the condition where the $\rm log \xi$ from \texttt{pion} is equal to \texttt{tpho}.
The right panels show $\rm log \xi$ versus $ C_{\rm Fe}$, where the blue dots represent the \texttt{pion} $\rm log \xi$ and red dots represent the \texttt{tpho} $\rm log \xi$ respectively.
}
\label{fig:pion_tpho}
\end{figure*}
%============================

In our current work, we derive the density of ionized plasma using the \texttt{tpho} model.
In \texttt{tpho} calculations, the $\rm \xi$ value simply scales with the time-dependent luminosity value, assuming that the product $n_{\rm H} \times r^2$ remains constant over time. 
Thus, 
\begin{equation}\label{equ: TPHO xi}
    \xi_{\rm TPHO}(t) \equiv  \frac{L(t)}{L_{\circ}} \times \xi_{\circ} \, ,
\end{equation}
where $L_{\circ}$ and $\xi_{\circ}$ are the luminosity and ionization parameter at $t = 0$, which is assumed to be in equilibrium state.

The \texttt{pion} model assumes photoionization equilibrium, and the ionization parameter $\rm \xi$ is obtained by fitting the observed spectra.
Once the $\rm \xi$ value is known, the ion concentrations of different elements are known and the corresponding elemental average charge is also determined.
So there is a one-to-one relation of $\xi_{\rm PION}$ to the element average charge in the \texttt{pion} equilibrium state.

In the left panels of Figure \ref{fig:pion_tpho}, the black dots represent the $\xi$ values of the \texttt{pion} model that are needed to yield the same Fe average charge value as the $C_{\rm Fe}(t)$ from the \texttt{tpho} simulation, represented by $\xi_{\rm TPHO}(t)$. 
The red line corresponds to equal values of both quantities. 
Only a small fraction of data points are located on the red line, which corresponds to the equilibrium state.

This figure also shows the limitations of applying an equilibrium (\texttt{pion}) model to a source that in reality has time-dependent photoionization, as represented by the realistic \texttt{tpho} model.
This holds in particular for periods of low luminosity (low $\xi_{\rm TPHO}(t)$).
In the right panels of Fig. \ref{fig:pion_tpho}, for example, $n_{\rm H} = 10^{12} \, \rm m^{-3}$, the data point with the lowest $\xi_{\rm TPHO}$ value of $2.75$ (in the log) has an average charge for Fe of $25.4$; if the corresponding spectrum would be fitted with an equilibrium \texttt{pion} model, this average charge would yield $\rm log \xi = 3.5$, a $\xi$ value that is $5.6$ times larger than the \texttt{tpho} value.
Because the observer would like to use the measured luminosity for this data point to derive the product $n_{\rm H} \times r^2 = L_{\rm ion}/\xi$, both the lower $L$ and larger $\xi$ would lead to an underestimate of $n_{\rm H} \times r^2$ by more than an order of magnitude.

%----------------------------------------------------------------------------
\section{Discussion}\label{sect:4}

\subsection{NGC $3783$ $10$ WAs geometry distance}
\begin{figure}[!tbp]
\centering
\includegraphics[width=0.9\linewidth]{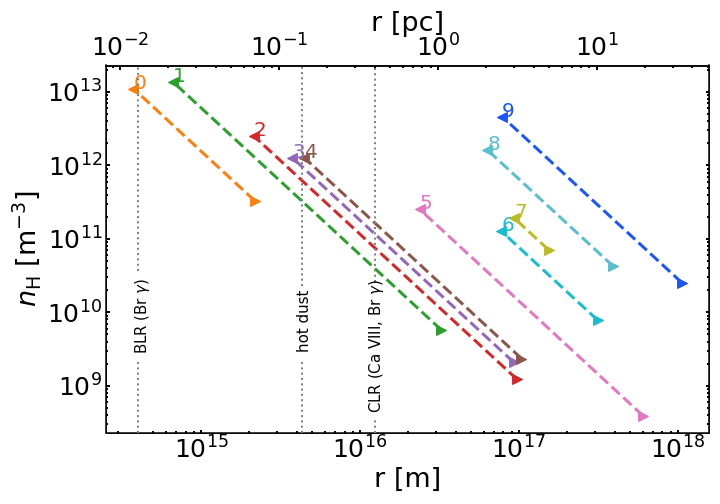}
\caption{ NGC $3783$ $10$ WAs predict detectable distance range with \texttt{tpho}-delay method in colored dash line.
The vertical dash line are main components of structure measured by \cite{GRAVITYCollaboration2021A&A}.
}
\label{fig:nH_distance}
\end{figure}

We use the Eq. \ref{equ:xi defination} to derive the detectable distance based on our lag density relation in current work and compare it with the NGC $3783$ intrinsic structure measured by \cite{GRAVITYCollaboration2021A&A} as shown in Fig. \ref{fig:nH_distance}.
Component $0$ exhibits a detectable distance very close to the broad line region, while components $1-2$ can be constrained within the host dust structure. Component $1-4$ shows a relatively large detectable distance range, typically within or around $1$ parsec, which corresponds to the torus structure. Additionally, the distances of components $5-9$ are detectable with our method only when they locate outside the torus.

%============================
%  Fig: density vs. xil
%
%============================

\subsection{Correspondence with the UV-X-ray absorbers}

\begin{figure}[!tbp]
\centering
\includegraphics[width=\linewidth]{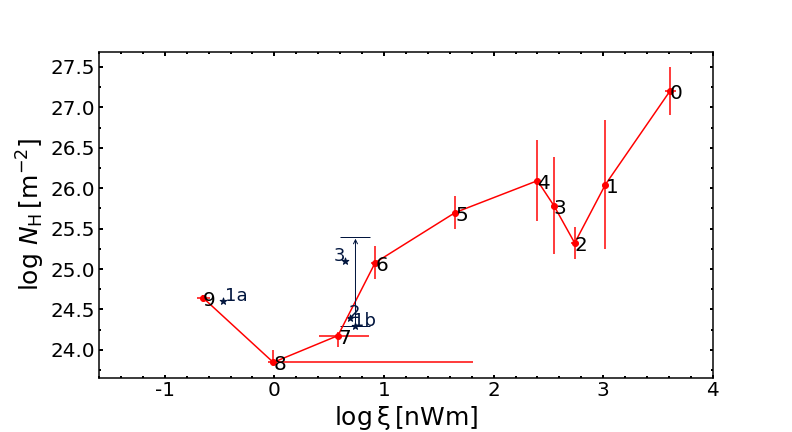}
\includegraphics[width=\linewidth]{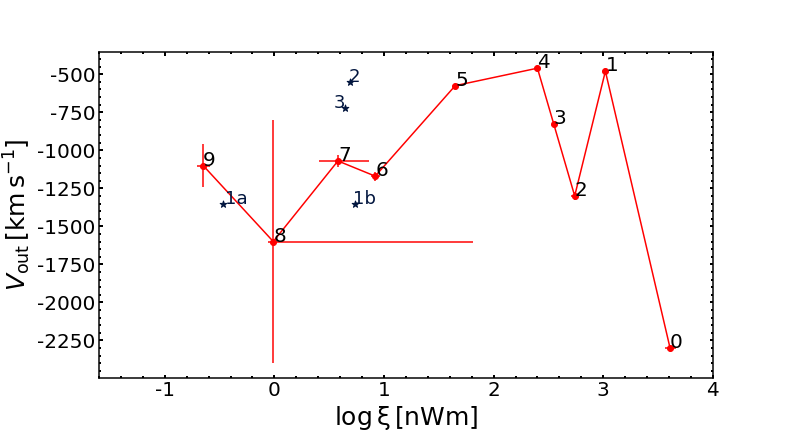}
\caption{
The $10$ X-ray WA components and $4$ UV components (\citealp{Gabel2005ApJ}) in NGC $3783$ observations.
Top panel: $ N_{\rm H}$ versus $\rm log \, \xi$. 
Bottom panel: $V_{\rm out}$ versus $\rm log \xi$.
}
\label{fig:NH_xil and vout_xil}
\end{figure}

In addition to the $10$ X-ray WAs, there are absorbers detected in UV (\citealp{Kraemer2001ApJ}; \citealp{Gabel2005ApJ}) that exhibit a bit of common features as the chosen X-ray counterpart.
Figure \ref{fig:summary methods} shows three UV absorbers reported in \cite{Gabel2005ApJ}, in which the component $1$ is composed of two physically distinct regions ($1a$ and $1b$) detected by the difference in covering factor and kinematic structure.
It denotes that both components $1a$ and $1b$ are assumed that they are colocated based on a decrease in radial velocities of lines in both components together with the same outflowing velocity (bottom panel of Fig. \ref{fig:NH_xil and vout_xil}), while the ionization parameter for component $1b$ derived in the unsaturated red wing of \ion{C}{IV} and \ion{N}{V} is higher than that of $1a$, and therefore component $1b$ has lower density and column density (top panel of Fig. \ref{fig:NH_xil and vout_xil}). 
The electron density of the lower ionization component $1a$, $\sim 3\times 10^{10} \, \rm m^{-3}$, obtained directly from the individual meta-stable levels, falls in the range where the \texttt{tpho}-delay method, as well as the metastable lines of \ion{Ne}{V}, can provide useful constraints.
Moreover, UV component $1a$ appears to be intimately linked to the component $9$ in the X-ray.
Therefore, UV component $1a$ can serve as a calibration for density diagnostic using an X-ray detector.

UV components $2$ and $3$ do not have direct estimates of the density and thus are less well constrained than component $1$ (Fig. \ref{fig:summary methods}). 
The lower limits on their densities are derived based on the basis of the observed variability. 
The ionization parameters from higher ionization UV absorbers, components $1b$, $2$, and $3$ with $\rm log \xi \sim 0.7$, are close to the X-ray component $7$ with $\rm log \xi \sim 0.6$.
Nevertheless, the latter has a smaller total column density (Fig. \ref{fig:NH_xil and vout_xil}).

%============================
%  Fig: density vs. xil
%
%============================

Although the hints mentioned above suggest a possible connection between the UV absorbers and the lowly ionized X-ray counterparts, it is not yet established due to poor X-ray constraints.
This hypothesis can be addressed through the high-resolution X-ray data that will be obtained with the upcoming XRISM \citep{XRISM2022arXiv220205399X} and Athena missions \citep{Nandra2013arXiv1306.2307N}.

\subsection{Crossing time}
X-ray variability from black holes is generally characterized as a red noise process \citep{Uttley2014A&ARv}, which can be contributed by different physical radiation 
processes \citep{Edward2021iScience}. 
\cite{Gabel2003ApJ} note that the observed changes in the absorption components could be due to the transverse motion of the clumping wind that moves across the line of sight, except for the intrinsic luminosity variability.
To determine the potential impact of Keplerian motions on WA variability, we have calculated the crossing time of outflows and compared them with the recombination time.

The X-ray emitting region has a scale no larger than $20$ gravitational radii \citep{Reis2013ApJL} as shown in the following equation,
\begin{equation}\label{equa:X-ray regin}
    s = \frac{20 * G * M_{\rm BH}}{c^2} \, \, . 
\end{equation}
Therefore the crossing time $t_{\rm cross}$ can be computed,
\begin{equation}\label{equa:crossing time}
    t_{\rm cross} = \frac{s}{V_{\rm kepler}} \, \, ,
\end{equation}
where $V_{\rm kepler} = \sqrt{\frac{G \times M_{\rm BH}}{R_{\rm WA}}}$. 
Substituting Eq. \ref{equ:xi defination} and Eq. \ref{equa:X-ray regin} into Eq. \ref{equa:crossing time}, we come to,
\begin{equation}\label{equa:}
    t_{\rm cross} = \frac{20}{c^2} \, (G \times M_{\rm BH})^{1/2} \, (\frac{L_{\circ}}{\xi_{\circ}})^{1/4} \, n_{\rm H}^{-1/4}  \, \, .
\end{equation}
Here $L_{\circ} = 6.36 \times 10^{36} \, \rm W$ (\citealp{Mehdipour2017A&A}), $\xi_{\circ}$ from Table \ref{table:pion good-fit par}.

\begin{figure*}[!tbp]
\centering
\hspace*{-0.15cm}\resizebox{0.35\hsize}{!}{\includegraphics[angle=0]{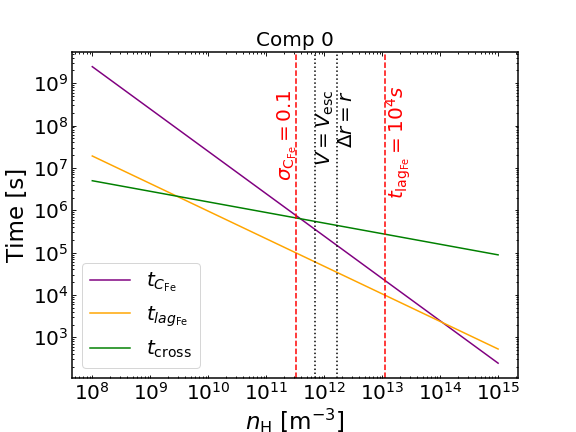}}
\hspace*{-0.15cm}\resizebox{0.35\hsize}{!}{\includegraphics[angle=0]{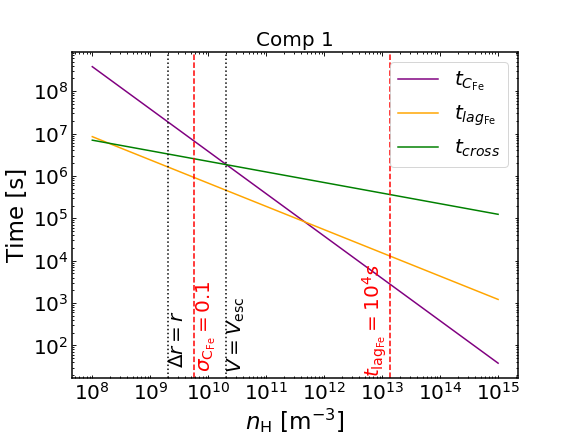}}
\hspace*{-0.15cm}\resizebox{0.35\hsize}{!}{\includegraphics[angle=0]{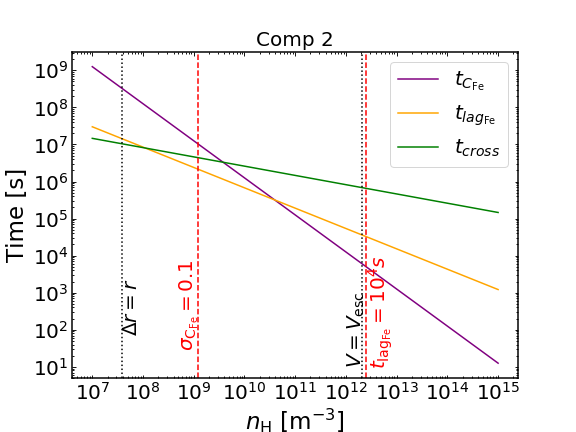}}
\hspace*{-0.15cm}\resizebox{0.35\hsize}{!}{\includegraphics[angle=0]{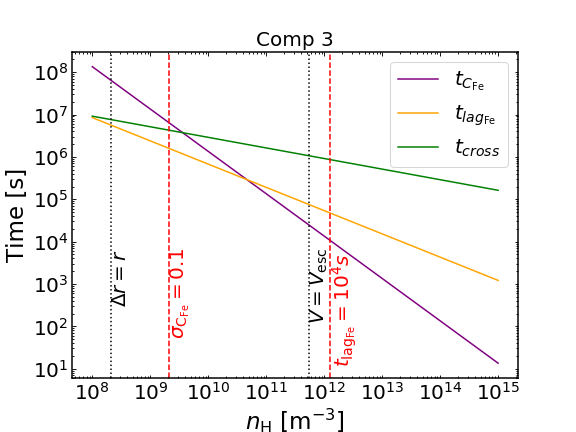}}
\hspace*{-0.15cm}\resizebox{0.35\hsize}{!}{\includegraphics[angle=0]{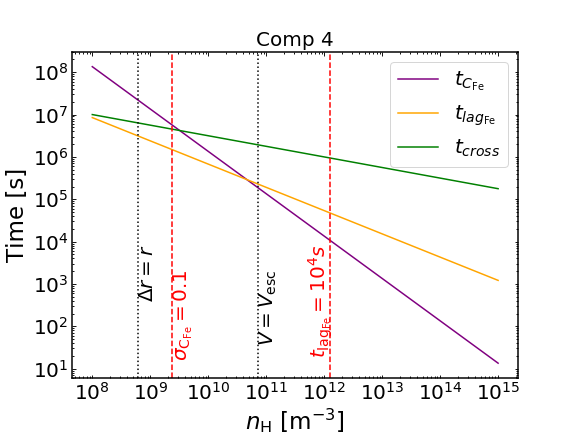}}
\hspace*{-0.15cm}\resizebox{0.35\hsize}{!}{\includegraphics[angle=0]{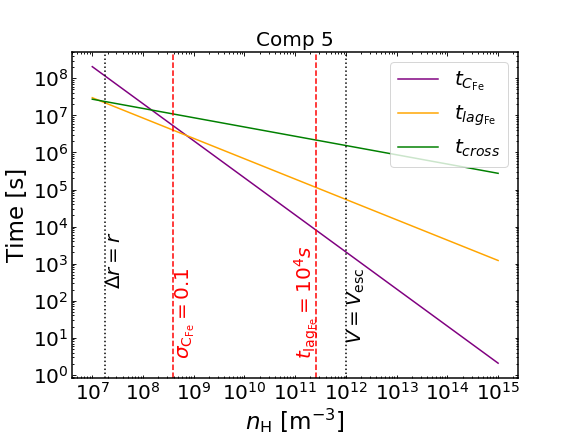}}
\hspace*{-0.15cm}\resizebox{0.35\hsize}{!}{\includegraphics[angle=0]{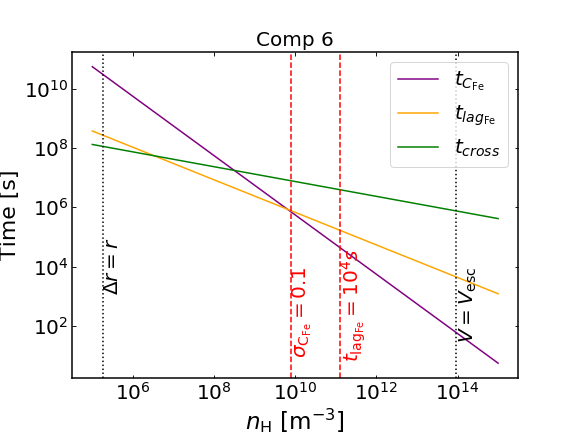}}
\hspace*{-0.15cm}\resizebox{0.35\hsize}{!}{\includegraphics[angle=0]{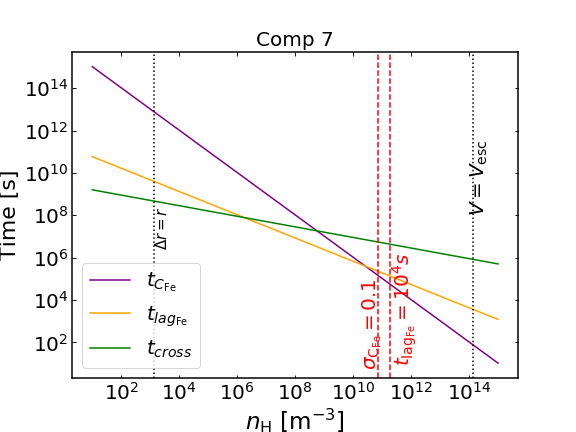}}
\hspace*{-0.15cm}\resizebox{0.35\hsize}{!}{\includegraphics[angle=0]{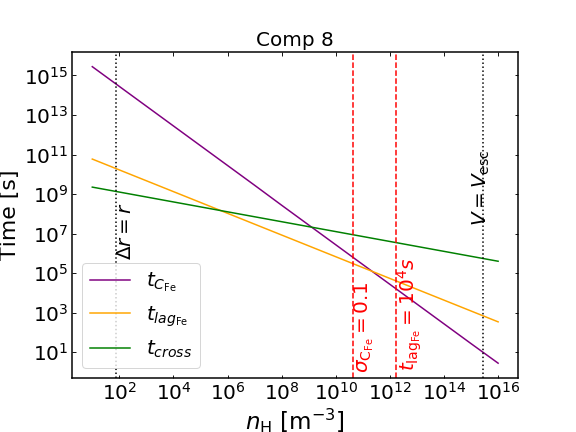}}
\hspace*{-0.15cm}\resizebox{0.35\hsize}{!}{\includegraphics[angle=0]{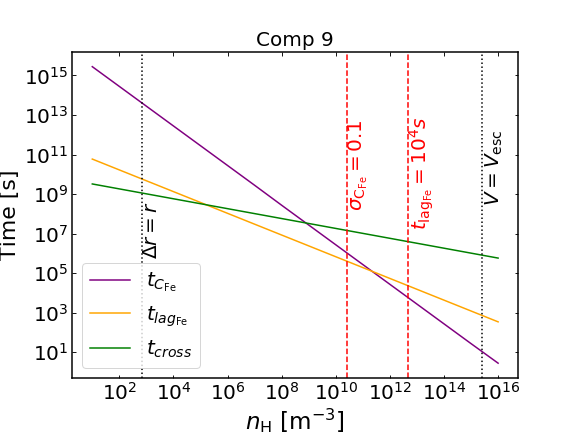}}
\caption{
Comparing $t_{C_{\rm Fe}}$, $t_{lag_{\rm Fe}}$ and $t_{\rm cross}$ of NGC 3783 10 X-ray WAs.
The dash and dot lines notations same as Fig.\ref{fig:summary methods}.
}
\label{fig:tcross}
\end{figure*}

The three timescales in Fig. \ref{fig:tcross}, charge time $t_{C_{\rm Fe}}$, lag timescale $t_{lag_{\rm Fe}}$, crossing time $t_{\rm cross}$, are density-dependent.
We also present the positions of density constraints from \texttt{tpho}-delay and other methods.

It shows that the density ranges where our \texttt{tpho}-delay method can offer useful constraints likely satisfy $t_{C_{\rm Fe}} < t_{\rm cross}$.
This indicates that the transverse motion of the outflows could not have a significant impact on the density diagnostics obtained assuming luminosity variation, i.e. the ionization variation is mainly/likely caused by the variability, not the transverse motions.
Hence, we consider the \texttt{tpho}-delay methods to be largely valid, even when accounting for the potential variability driven by motion.

%============================
%  Fig: density vs. xil
%
%============================
%\subsection{Uncertainties}
%ompare the transmission spectra between pion and tpho.

\subsection{SED and light curve of NGC 3783}

The AGN SEDs do not normally change their shape a lot unless something dramatic happens: like in changing-look AGN where the accretion rate and the X-ray corona emission decline. 
Or if there is strong X-ray obscuration, and the difference between the impact of obscured and unobscured SED on the ionization state is significant, as shown in NGC $5548$ \citep{Mehdipour2016A&A}.
Here, the SED we used is uniformly derived from an unobscured state to study the $10$ WA components ionization state.
Because the component $0$ is in the obscured state but not shielded by the obscurer, it is reasonable to assume that the same SED can be used for component $0$ and components $1-9$.
NGC $3783$ is a Seyfert-$1$ galaxy without extreme changing-look accretion events observed until now. 
So it is reasonable to say that it is the normalization that is mostly changing in an unobscured state. 
Of course, individual SED components (like the soft excess) show some special variability, but usually this does not cause the shape to change too much such that it would impact photoionization. 
Thus for \texttt{tpho} to assume that only the normalization of the SED is changing is a good enough approximation in this case. 
We also note that deriving the variability of the SED shape (with limited data) has its challenges and assumptions which are highly model dependent.

\subsection{Comparison with other time-resolved analyses}

Time-dependent photoionization modelling for outflowing winds has been described in other papers, but with different model constructions.
\cite{Silva2016A&A} and \cite{Juranova2022MNRAS} carry out time-dependent photoionization modeling using pre-calculated runs from Cloudy, assuming equilibrium conditions.
The \texttt{tpho} model \cite{Rogantini2022ApJ} however takes into account non-equlibrium state conditions and tracks the time-dependency of the heating and cooling process according to the lightcurve of the ionizing SED.

%\textbf{The TPHO model of \cite{Rogantini2022ApJ} keeps the SED shape constant while the normalization of the luminosity is able to be changed by following the input light curve.
%$TPHO considers the cooling and heating process simultaneously, which performs a self-consistent calculation solving the full time-dependent ionization state for all ionic species.}

We point out that, without considering the non-equilibrium time-dependency of heating and cooling process, to some extent, it will affect the resulting calculation of ion concentration, because recombination rates are not only a function of density but also temperature.
%The impact of temperature on the reaction timescale is much larger than density somewhere.
A long light curve history is required for \texttt{tpho} input, and the upcoming new satellite XRISM can provide good spectra to compare with the models. 
It is expected that \texttt{tpho} can be used to fit the observational data in the case that AGNs SED do not change their shape too much.

\section{Conclusions}\label{sect:5}

By utilizing a realistic SED with variability from a realistic PSD, and WA physical parameters based on previous X-ray observations, we performed a theoretical \texttt{tpho} model calculation to investigate the time-dependent photoionization plasma state of 10 WA components of NGC 3783. 
We have employed cross-correlation calculations to measure the correlation between response lag time and the outflow density for all WA components.
By doing so, we have determined the ranges of density for which the \texttt{tpho} model can yield practical constraints, and we have compared our results with those obtained from metastable absorptions lines in X-ray and UV bands.
We further prove that the Keplerian motion would have negligible impact on the density measurement for our range of interest.
It is expected that our technique will be applied to the new data obtained with upcoming X-ray missions including XRISM and Athena.

%
%                                                One column figure
%----------------------------------------------------------------- 

%-----------------------------------------------------------------

\begin{acknowledgements}
      We thank the anonymous referee for his/her constructive comments.
      C.L. acknowledges support from Chinese Scholarship Council (CSC) and Leiden University/Leiden Observatory, as well as SRON.
      SRON is supported financially by NWO, the Netherlands Organization for Scientific Research.
      C.L. thanks Anna Jur\'{a}\u{n}ov\'{a} for the discussions of different time-dependent photoionization model constructions.
      %We thank the anonymous referee for his/her constructive comments.
\end{acknowledgements}


%\begin{thebibliography}{}
%\end{thebibliography}
\bibliographystyle{aa} % style aa.bst
\bibliography{Li_NGC3783.bib}

\begin{thebibliography}{36}
\expandafter\ifx\csname natexlab\endcsname\relax\def\natexlab#1{#1}\fi

\bibitem[{{Arav} {et~al.}(2015){Arav}, {Chamberlain}, {Kriss}, {Kaastra},
  {Cappi}, {Mehdipour}, {Petrucci}, {Steenbrugge}, {Behar}, {Bianchi},
  {Boissay}, {Branduardi-Raymont}, {Costantini}, {Ely}, {Ebrero}, {di Gesu},
  {Harrison}, {Kaspi}, {Malzac}, {De Marco}, {Matt}, {Nandra}, {Paltani},
  {Peterson}, {Pinto}, {Ponti}, {Pozo Nu{\~n}ez}, {De Rosa}, {Seta}, {Ursini},
  {de Vries}, {Walton}, \& {Whewell}}]{Arav2015AA}
{Arav}, N., {Chamberlain}, C., {Kriss}, G.~A., {et~al.} 2015, \aap, 577, A37

\bibitem[{{Blustin} {et~al.}(2005){Blustin}, {Page}, {Fuerst},
  {Branduardi-Raymont}, \& {Ashton}}]{Blustin2005A&A}
{Blustin}, A.~J., {Page}, M.~J., {Fuerst}, S.~V., {Branduardi-Raymont}, G., \&
  {Ashton}, C.~E. 2005, \aap, 431, 111

\bibitem[{{Bottorff} {et~al.}(2000){Bottorff}, {Korista}, \&
  {Shlosman}}]{Bottorff2000ApJ}
{Bottorff}, M.~C., {Korista}, K.~T., \& {Shlosman}, I. 2000, \apj, 537, 134

\bibitem[{Cackett {et~al.}(2021)Cackett, Bentz, \& Kara}]{Edward2021iScience}
Cackett, E.~M., Bentz, M.~C., \& Kara, E. 2021, iScience, 24, 102557

\bibitem[{{Davies} {et~al.}(2015){Davies}, {Burtscher}, {Rosario},
  {Storchi-Bergmann}, {Contursi}, {Genzel}, {Graci{\'a}-Carpio}, {Hicks},
  {Janssen}, {Koss}, {Lin}, {Lutz}, {Maciejewski}, {M{\"u}ller-S{\'a}nchez},
  {Orban de Xivry}, {Ricci}, {Riffel}, {Riffel}, {Schartmann},
  {Schnorr-M{\"u}ller}, {Sternberg}, {Sturm}, {Tacconi}, \&
  {Veilleux}}]{Davies2015ApJ}
{Davies}, R.~I., {Burtscher}, L., {Rosario}, D., {et~al.} 2015, \apj, 806, 127

\bibitem[{{Ebrero} {et~al.}(2016){Ebrero}, {Kaastra}, {Kriss}, {Di Gesu},
  {Costantini}, {Mehdipour}, {Bianchi}, {Cappi}, {Boissay},
  {Branduardi-Raymont}, {Petrucci}, {Ponti}, {Pozo N{\'u}{\~n}ez}, {Seta},
  {Steenbrugge}, \& {Whewell}}]{Ebrero2016A&A}
{Ebrero}, J., {Kaastra}, J.~S., {Kriss}, G.~A., {et~al.} 2016, \aap, 587, A129

\bibitem[{{Ferland} {et~al.}(1998){Ferland}, {Korista}, {Verner}, {Ferguson},
  {Kingdon}, \& {Verner}}]{Ferland1998PASP}
{Ferland}, G.~J., {Korista}, K.~T., {Verner}, D.~A., {et~al.} 1998, \pasp, 110,
  761

\bibitem[{{Gabel} {et~al.}(2003){Gabel}, {Crenshaw}, {Kraemer}, {Brandt},
  {George}, {Hamann}, {Kaiser}, {Kaspi}, {Kriss}, {Mathur}, {Mushotzky},
  {Nandra}, {Netzer}, {Peterson}, {Shields}, {Turner}, \&
  {Zheng}}]{Gabel2003ApJ}
{Gabel}, J.~R., {Crenshaw}, D.~M., {Kraemer}, S.~B., {et~al.} 2003, \apj, 595,
  120

\bibitem[{{Gabel} {et~al.}(2005){Gabel}, {Kraemer}, {Crenshaw}, {George},
  {Brandt}, {Hamann}, {Kaiser}, {Kaspi}, {Kriss}, {Mathur}, {Nandra}, {Netzer},
  {Peterson}, {Shields}, {Turner}, \& {Zheng}}]{Gabel2005ApJ}
{Gabel}, J.~R., {Kraemer}, S.~B., {Crenshaw}, D.~M., {et~al.} 2005, \apj, 631,
  741

\bibitem[{{GRAVITY Collaboration} {et~al.}(2021){GRAVITY Collaboration},
  {Amorim}, {Baub{\"o}ck}, {Brandner}, {Bolzer}, {Cl{\'e}net}, {Davies}, {de
  Zeeuw}, {Dexter}, {Drescher}, {Eckart}, {Eisenhauer}, {F{\"o}rster
  Schreiber}, {Gao}, {Garcia}, {Genzel}, {Gillessen}, {Gratadour}, {H{\"o}nig},
  {Kaltenbrunner}, {Kishimoto}, {Lacour}, {Lutz}, {Millour}, {Netzer}, {Ott},
  {Paumard}, {Perraut}, {Perrin}, {Peterson}, {Petrucci}, {Pfuhl}, {Prieto},
  {Rouan}, {Sanchez-Bermudez}, {Shangguan}, {Shimizu}, {Schartmann}, {Stadler},
  {Sternberg}, {Straub}, {Straubmeier}, {Sturm}, {Tacconi}, {Tristram},
  {Vermot}, {von Fellenberg}, {Waisberg}, {Widmann}, \&
  {Woillez}}]{GRAVITYCollaboration2021A&A}
{GRAVITY Collaboration}, {Amorim}, A., {Baub{\"o}ck}, M., {et~al.} 2021, \aap,
  648, A117

\bibitem[{{Jur{\'a}{\v{n}}ov{\'a}} {et~al.}(2022){Jur{\'a}{\v{n}}ov{\'a}},
  {Costantini}, \& {Uttley}}]{Juranova2022MNRAS}
{Jur{\'a}{\v{n}}ov{\'a}}, A., {Costantini}, E., \& {Uttley}, P. 2022, \mnras,
  510, 4225

\bibitem[{{Kaastra} {et~al.}(2012){Kaastra}, {Detmers}, {Mehdipour}, {Arav},
  {Behar}, {Bianchi}, {Branduardi-Raymont}, {Cappi}, {Costantini}, {Ebrero},
  {Kriss}, {Paltani}, {Petrucci}, {Pinto}, {Ponti}, {Steenbrugge}, \& {de
  Vries}}]{Kaastra2012A&A}
{Kaastra}, J.~S., {Detmers}, R.~G., {Mehdipour}, M., {et~al.} 2012, \aap, 539,
  A117

\bibitem[{{Kaastra} {et~al.}(2022){Kaastra}, {Raassen}, {de Plaa}, \&
  {Gu}}]{kaastra2022}
{Kaastra}, J.~S., {Raassen}, A.~J.~J., {de Plaa}, J., \& {Gu}, L. 2022, {SPEX
  X-ray spectral fitting package}, Zenodo

\bibitem[{{Kaspi} {et~al.}(2002){Kaspi}, {Brandt}, {George}, {Netzer},
  {Crenshaw}, {Gabel}, {Hamann}, {Kaiser}, {Koratkar}, {Kraemer}, {Kriss},
  {Mathur}, {Mushotzky}, {Nandra}, {Peterson}, {Shields}, {Turner}, \&
  {Zheng}}]{Kaspi2002ApJ}
{Kaspi}, S., {Brandt}, W.~N., {George}, I.~M., {et~al.} 2002, \apj, 574, 643

\bibitem[{{Kraemer} {et~al.}(2001){Kraemer}, {Crenshaw}, \&
  {Gabel}}]{Kraemer2001ApJ}
{Kraemer}, S.~B., {Crenshaw}, D.~M., \& {Gabel}, J.~R. 2001, \apj, 557, 30

\bibitem[{{Krolik} \& {Kriss}(1995)}]{Krolik&Kriss1995ApJ}
{Krolik}, J.~H. \& {Kriss}, G.~A. 1995, \apj, 447, 512

\bibitem[{{Krolik} \& {Kriss}(2001)}]{Krolik2001ApJ}
{Krolik}, J.~H. \& {Kriss}, G.~A. 2001, \apj, 561, 684

\bibitem[{{Krolik} {et~al.}(1981){Krolik}, {McKee}, \&
  {Tarter}}]{Krolik1981ApJ}
{Krolik}, J.~H., {McKee}, C.~F., \& {Tarter}, C.~B. 1981, \apj, 249, 422

\bibitem[{{Laha} {et~al.}(2021){Laha}, {Reynolds}, {Reeves}, {Kriss},
  {Guainazzi}, {Smith}, {Veilleux}, \& {Proga}}]{Laha2021NatAs}
{Laha}, S., {Reynolds}, C.~S., {Reeves}, J., {et~al.} 2021, Nature Astronomy,
  5, 13

\bibitem[{{Mao} {et~al.}(2017){Mao}, {Kaastra}, {Mehdipour}, {Raassen}, {Gu},
  \& {Miller}}]{Mao2017A&A}
{Mao}, J., {Kaastra}, J.~S., {Mehdipour}, M., {et~al.} 2017, \aap, 607, A100

\bibitem[{{Mao} {et~al.}(2019){Mao}, {Mehdipour}, {Kaastra}, {Costantini},
  {Pinto}, {Branduardi-Raymont}, {Behar}, {Peretz}, {Bianchi}, {Kriss},
  {Ponti}, {De Marco}, {Petrucci}, {Di Gesu}, {Middei}, {Ebrero}, \&
  {Arav}}]{Mao2019A&A}
{Mao}, J., {Mehdipour}, M., {Kaastra}, J.~S., {et~al.} 2019, \aap, 621, A99

\bibitem[{{Markowitz}(2005)}]{Markowitz2005ApJ}
{Markowitz}, A. 2005, \apj, 635, 180

\bibitem[{{Mehdipour} {et~al.}(2016){Mehdipour}, {Kaastra}, \&
  {Kallman}}]{Mehdipour2016A&A}
{Mehdipour}, M., {Kaastra}, J.~S., \& {Kallman}, T. 2016, \aap, 596, A65

\bibitem[{{Mehdipour} {et~al.}(2017){Mehdipour}, {Kaastra}, {Kriss}, {Arav},
  {Behar}, {Bianchi}, {Branduardi-Raymont}, {Cappi}, {Costantini}, {Ebrero},
  {Di Gesu}, {Kaspi}, {Mao}, {De Marco}, {Matt}, {Paltani}, {Peretz},
  {Peterson}, {Petrucci}, {Pinto}, {Ponti}, {Ursini}, {de Vries}, \&
  {Walton}}]{Mehdipour2017A&A}
{Mehdipour}, M., {Kaastra}, J.~S., {Kriss}, G.~A., {et~al.} 2017, \aap, 607,
  A28

\bibitem[{{Nandra} {et~al.}(2013){Nandra}, {Barret}, {Barcons}, {Fabian}, {den
  Herder}, {Piro}, {Watson}, {Adami}, {Aird}, {Afonso}, {Alexander},
  {Argiroffi}, {Amati}, {Arnaud}, {Atteia}, {Audard}, {Badenes}, {Ballet},
  {Ballo}, {Bamba}, {Bhardwaj}, {Stefano Battistelli}, {Becker}, {De Becker},
  {Behar}, {Bianchi}, {Biffi}, {B{\^\i}rzan}, {Bocchino}, {Bogdanov}, {Boirin},
  {Boller}, {Borgani}, {Borm}, {Bouch{\'e}}, {Bourdin}, {Bower}, {Braito},
  {Branchini}, {Branduardi-Raymont}, {Bregman}, {Brenneman}, {Brightman},
  {Br{\"u}ggen}, {Buchner}, {Bulbul}, {Brusa}, {Bursa}, {Caccianiga},
  {Cackett}, {Campana}, {Cappelluti}, {Cappi}, {Carrera}, {Ceballos},
  {Christensen}, {Chu}, {Churazov}, {Clerc}, {Corbel}, {Corral}, {Comastri},
  {Costantini}, {Croston}, {Dadina}, {D'Ai}, {Decourchelle}, {Della Ceca},
  {Dennerl}, {Dolag}, {Done}, {Dovciak}, {Drake}, {Eckert}, {Edge}, {Ettori},
  {Ezoe}, {Feigelson}, {Fender}, {Feruglio}, {Finoguenov}, {Fiore}, {Galeazzi},
  {Gallagher}, {Gandhi}, {Gaspari}, {Gastaldello}, {Georgakakis},
  {Georgantopoulos}, {Gilfanov}, {Gitti}, {Gladstone}, {Goosmann}, {Gosset},
  {Grosso}, {Guedel}, {Guerrero}, {Haberl}, {Hardcastle}, {Heinz}, {Alonso
  Herrero}, {Herv{\'e}}, {Holmstrom}, {Iwasawa}, {Jonker}, {Kaastra}, {Kara},
  {Karas}, {Kastner}, {King}, {Kosenko}, {Koutroumpa}, {Kraft}, {Kreykenbohm},
  {Lallement}, {Lanzuisi}, {Lee}, {Lemoine-Goumard}, {Lobban}, {Lodato},
  {Lovisari}, {Lotti}, {McCharthy}, {McNamara}, {Maggio}, {Maiolino}, {De
  Marco}, {de Martino}, {Mateos}, {Matt}, {Maughan}, {Mazzotta}, {Mendez},
  {Merloni}, {Micela}, {Miceli}, {Mignani}, {Miller}, {Miniutti}, {Molendi},
  {Montez}, {Moretti}, {Motch}, {Naz{\'e}}, {Nevalainen}, {Nicastro}, {Nulsen},
  {Ohashi}, {O'Brien}, {Osborne}, {Oskinova}, {Pacaud}, {Paerels}, {Page},
  {Papadakis}, {Pareschi}, {Petre}, {Petrucci}, {Piconcelli}, {Pillitteri},
  {Pinto}, {de Plaa}, {Pointecouteau}, {Ponman}, {Ponti}, {Porquet}, {Pounds},
  {Pratt}, {Predehl}, {Proga}, {Psaltis}, {Rafferty}, {Ramos-Ceja}, {Ranalli},
  {Rasia}, {Rau}, {Rauw}, {Rea}, {Read}, {Reeves}, {Reiprich}, {Renaud},
  {Reynolds}, {Risaliti}, {Rodriguez}, {Rodriguez Hidalgo}, {Roncarelli},
  {Rosario}, {Rossetti}, {Rozanska}, {Rovilos}, {Salvaterra}, {Salvato}, {Di
  Salvo}, {Sanders}, {Sanz-Forcada}, {Schawinski}, {Schaye}, {Schwope},
  {Sciortino}, {Severgnini}, {Shankar}, {Sijacki}, {Sim}, {Schmid}, {Smith},
  {Steiner}, {Stelzer}, {Stewart}, {Strohmayer}, {Str{\"u}der}, {Sun}, {Takei},
  {Tatischeff}, {Tiengo}, {Tombesi}, {Trinchieri}, {Tsuru}, {Ud-Doula},
  {Ursino}, {Valencic}, {Vanzella}, {Vaughan}, {Vignali}, {Vink}, {Vito},
  {Volonteri}, {Wang}, {Webb}, {Willingale}, {Wilms}, {Wise}, {Worrall},
  {Young}, {Zampieri}, {In't Zand}, {Zane}, {Zezas}, {Zhang}, \&
  {Zhuravleva}}]{Nandra2013arXiv1306.2307N}
{Nandra}, K., {Barret}, D., {Barcons}, X., {et~al.} 2013, arXiv e-prints,
  arXiv:1306.2307

\bibitem[{{Nicastro} {et~al.}(1999){Nicastro}, {Fiore}, {Perola}, \&
  {Elvis}}]{Nicastro1999ApJ}
{Nicastro}, F., {Fiore}, F., {Perola}, G.~C., \& {Elvis}, M. 1999, \apj, 512,
  184

\bibitem[{{Porquet} {et~al.}(2010){Porquet}, {Dubau}, \&
  {Grosso}}]{Porquet2010SSR}
{Porquet}, D., {Dubau}, J., \& {Grosso}, N. 2010, \ssr, 157, 103

\bibitem[{{Reis} \& {Miller}(2013)}]{Reis2013ApJL}
{Reis}, R.~C. \& {Miller}, J.~M. 2013, \apjl, 769, L7

\bibitem[{{Rogantini} {et~al.}(2022){Rogantini}, {Mehdipour}, {Kaastra},
  {Costantini}, {Jur{\'a}{\v{n}}ov{\'a}}, \& {Kara}}]{Rogantini2022ApJ}
{Rogantini}, D., {Mehdipour}, M., {Kaastra}, J., {et~al.} 2022, \apj, 940, 122

\bibitem[{{Sadaula} {et~al.}(2022){Sadaula}, {Bautista}, {Garcia}, \&
  {Kallman}}]{Sadaula2022arXiv}
{Sadaula}, D.~R., {Bautista}, M.~A., {Garcia}, J.~A., \& {Kallman}, T.~R. 2022,
  arXiv e-prints, arXiv:2205.04708

\bibitem[{{Silva} {et~al.}(2016){Silva}, {Uttley}, \&
  {Costantini}}]{Silva2016A&A}
{Silva}, C.~V., {Uttley}, P., \& {Costantini}, E. 2016, \aap, 596, A79

\bibitem[{{Tarter} {et~al.}(1969){Tarter}, {Tucker}, \&
  {Salpeter}}]{Tarter1969ApJ}
{Tarter}, C.~B., {Tucker}, W.~H., \& {Salpeter}, E.~E. 1969, \apj, 156, 943

\bibitem[{{Theureau} {et~al.}(1998){Theureau}, {Bottinelli}, {Coudreau-Durand},
  {Gouguenheim}, {Hallet}, {Loulergue}, {Paturel}, \&
  {Teerikorpi}}]{Theureau1998A&AS}
{Theureau}, G., {Bottinelli}, L., {Coudreau-Durand}, N., {et~al.} 1998, \aaps,
  130, 333

\bibitem[{{Uttley} {et~al.}(2014){Uttley}, {Cackett}, {Fabian}, {Kara}, \&
  {Wilkins}}]{Uttley2014A&ARv}
{Uttley}, P., {Cackett}, E.~M., {Fabian}, A.~C., {Kara}, E., \& {Wilkins},
  D.~R. 2014, \aapr, 22, 72

\bibitem[{{Vestergaard} \& {Peterson}(2006)}]{Vestergaard2006ApJ}
{Vestergaard}, M. \& {Peterson}, B.~M. 2006, \apj, 641, 689

\bibitem[{{XRISM Science Team}(2022)}]{XRISM2022arXiv220205399X}
{XRISM Science Team}. 2022, arXiv e-prints, arXiv:2202.05399

\end{thebibliography}

%%%%%%%%%%%%%%%%%%%%%%%%%%%%%%%%%%%%%%%%%%%%%
% add appendix
%\appendix
%add table of tlagmin,tlagmax(CFe=0.1) and corresponding to density. 

\end{document}